\documentclass[useAMS,usenatbib]{mnras}
\usepackage{graphicx,verbatim}
\usepackage[normalem]{ulem}
\usepackage{color,ulem}
\usepackage{newtxtext}
\usepackage[T1]{fontenc} 
\usepackage{amsmath,amssymb}
\usepackage{lmodern}

\def \be{\begin{equation}}
\def \ee{\end{equation}}
\newcommand       \ba           {\begin{eqnarray}}
\newcommand       \ea           {\end{eqnarray}}
\def \bea{\begin{eqnarray}}
\def \eea{\end{eqnarray}}

\newcommand{\comments}[1]{}

\defcitealias{dha16}{Paper I}

\definecolor{webgreen}{rgb}{0,.5,0}
\definecolor{webbrown}{rgb}{.6,0,0}
\hypersetup{%
   colorlinks=true,hyperfootnotes=false,%
   breaklinks=true,%
   plainpages=false, bookmarksnumbered, bookmarksopen=true,
   bookmarksopenlevel=1,%
   urlcolor=webbrown, linkcolor=webbrown, citecolor=webgreen,
   }

\title[Magnetized SASI]{Magnetized SASI: its mechanism and possible connection to some QPOs in XRBs}   

\author[P. Dhang, P. Sharma, \textit{\&} B. Mukhopadhyay]
{Prasun Dhang,$^{1}$\thanks{E-mail:prasundhang@gmail.com}
Prateek Sharma,$^{1}$
Banibrata Mukhopadhyay$^{1}$
\\
$^{1}$Department of Physics and Joint Astronomy Programme, Indian Institute of Science, Bangalore, INDIA 560012
}

\begin{document}
\maketitle

\label{firstpage}

\begin{abstract}
 The presence of a surface at the inner boundary, such as in a neutron star or a white dwarf, allows the existence of a standing shock 
 in steady spherical accretion.
 The standing shock can become unstable in 2D or 3D; this is called the {\em standing accretion shock instability} (SASI).
 Two mechanisms -- advective-acoustic and purely acoustic -- have been proposed to explain SASI. Using axisymmetric hydrodynamic (HD) and magnetohydrodynamic (MHD)
 simulations, we find that the advective-acoustic mechanism better matches the observed oscillation timescales in our simulations. The global shock oscillations present 
 in the accretion flow can explain many observed high frequency ($\gtrsim 100$ Hz) quasi-periodic oscillations (QPOs) in X-ray binaries (XRBs). The presence of a moderately 
 strong magnetic field adds more features to the shock oscillation
 pattern, giving rise to low frequency  modulation in the computed light curve.  This low frequency modulation 
 can be responsible for  $\sim 100$ Hz QPOs (known as hHz QPOs). We propose that the appearance of hHz QPO determines the separation of twin peak QPOs of higher frequencies.
 
 \end{abstract}


% ...................................................................................................................Introduction.......................................................................................................................%

\section{Introduction}           %% first-level sections will be auto-capitalized
\label{sect:intro}
Spherically symmetric steady state accretion of adiabatic gas on to a point mass that can accrete the supersonically infalling gas (e.g., a black hole), 
is characterized by the classical transonic solution given by Bondi (\citealt{Bondi1952}).
 On the other hand, if the central accretor has a surface that accretes very slowly, a standing shock may form within the sonic radius (\citealt{McCrea1956}). For a detailed discussion, see 
section 5.1 of \citet{Dhang2016} (hereafter Paper I). 

The standing shock is stable in 1D under radial perturbations, but is unstable in 2D. The shock structure oscillates with
$l=1$ and higher order modes (axisymmetric sloshing modes). In the context of supernovae, \cite{Herant1994} advocated convective instability as the possible mechanism behind 
the oscillations of the stalled shock front. 
But, \cite{Foglizzo2006} showed that in presence of advection in the post-shock region, negative entropy gradient  is no longer a sufficient condition for
convective instability; advection acts as a stabilizing factor. %Even in the absence of any entropy gradient 
The shock instability exists even in the absence of an entropy gradient (\citealt{Blondin2003}, \citealt{Dhang2016}).
\cite{Blondin2003} named this instability as {\em standing accretion shock instability}, or SASI and identified advective-acoustic feedback (\citealt{Foglizzo2002}) as its possible
mechanism. Later, \cite{Blondin_mezzacappa2006} attributed SASI to a purely acoustic cycle, and thus triggering the debate on the physical origin of SASI.
Some other studies reached divergent conclusions. While studies of \cite{Ohnishi2006} and \cite{Scheck2008} identified advective-acoustic cycle as the possible mechanism, \cite{Laming2007}
in his analytical studies claimed that both advective-acoustic and purely acoustic cycles can be possible depending on the ratio of  the shock radius to the inner radius.

In 3D, in addition to these axisymmetric modes, SASI also shows a non-axisymmetric spiral mode ($m=1$) (\citealt{Blondin2007}). \cite{Fernandez2010} interpreted spiral modes
as the combination of two sloshing modes, whereas, \cite{Blondin_shaw2007} showed that sloshing modes can be constructed by combining two equal and opposite non-axisymmetric 
spiral modes. According to \cite{Kazeroni2016}, spiral modes dominate the dynamics of SASI only if the ratio of the initial shock radius to the neutron star radius exceeds a critical
value. Otherwise, dynamics is dominated by the sloshing mode. The actual mechanism behind the shock instability is still not fully understood. 

SASI has been studied extensively in the context of stellar collapse simulations over the years including different aspects of physics (e. g., neutrino transport, cooling, rotation, magnetic 
fields). There are state of the art realistic simulations (\citealt{Marek_Janka2009}, \citealt{Burrows2006}, \citealt{Bruenn2006} ) in which neutrino transport, self-gravity of stellar
gas, nuclear equation of state are considered. Also there are simplified planar toy models of SASI without any extra physics (\citealt{Foglizzo2009}, \citealt{Sato2009}). Models of 
SASI considering the angular momentum of the infalling gas are markedly different from models without angular momentum (\citealt{Blondin2007}). Spiral modes become more
prominent relative to sloshing modes in presence of rotation both in linear (\citealt{Yamasaki2008}) and in non-linear regime (\citealt{Iwakami2009}). \cite{Endeve2010} and
\cite{Endeve2012} explored the effects of a weak magnetic field in the absence and  presence of rotation both in axisymmetric and non-axisymmetric simulations. While axisymmetric 
models give magnetic field amplification of the order of 2, non-axisymmetric models provide an amplification of order 4. They also observe that magnetic field beyond a certain
strength stabilizes SASI.

As discussed earlier, different studies reached divergent conclusions by inspecting the linear properties of eigen modes, including the fundamental mode and its harmonics.
In this paper we study the physics of SASI in the non-linear regime using numerical simulations and try to shed some light on its 
mechanism by two different approaches: i) by changing the ratio of the shock radius to the inner radius in hydrodynamic (HD) simulations; ii) by changing the 
magnetic field strength in magnetohydrodynamic (MHD) simulations. 
If SASI is an outcome  of a meridional acoustic cycle, the weak magnetic field in the downstream region close to the shock should not affect the oscillation timescales.
On the other hand, a somewhat stronger magnetic field close to the center can affect the
%weak magnetic field can affects 
radial advective-acoustic cycle (\citealt{Guilet2010}).

 In Paper I, using our hydrodynamic axisymmetric simulations, we proposed that SASI in accretion flows 
may give rise to some of the quasi-periodic oscillations (QPOs) observed in the light curves of X-ray binaries. 
Most of the proposed QPO mechanisms are based on the physics of test particle motion (e.g. \citealt{Strohmayer1996}, \citealt{Miller1998},
\citealt{Stella1999},  \citealt{Kluzniak2002}, \citealt{Kluzniak2004}, \citealt{Mukhopadhyay2009}), which is not affected by pressure and magnetic fields.
However, for a particular model, the QPO frequencies obtained considering bulk motion significantly differ from the ones corresponding to free particles (\citealt{Blaes2007}).
Along with our model, there are few models (e.g. \citealt{Kato1980}, \citealt{Kato1990}, \citealt{Ipser1991}, \citealt{Wagoner2001}, \citealt{Yang1995}, \citealt{Ryu1995}, \citealt{MSC1996}, \citealt{Chakrabarti2000}, \citealt{Mukhopadhyay_QPO2003}) where bulk motion of  the flow is considered to explain the origin of QPOs. 

In reality, accreting matter around a compact object has angular momentum and is magnetized. As a first step,  here we incorporate
magnetic field and  explore the origin of QPOs appearing in the light curve due to  SASI in a magnetized accreting medium. 
This will help to understand the sole effect of magnetic field on SASI and QPOs.
Our particular emphasis is QPO frequencies $\gtrsim 100$ Hz in X-ray binaries, the origin of which is still not understood. We show that the presence of magnetic fields, hence magnetized SASI, appears to uncover some of the 
important characteristics of QPOs. In other words, the inclusion of magnetic fields introduces important
physics in the SASI model to predict certain QPOs, which is absent in a unmagnetized case.

The paper is organized as follows. In Section \ref{sect:mechanism}, we briefly discuss the two different mechanisms proposed to explain SASI.  In Section \ref{sect:method} we 
describe the physical set-up and the solution method. In Section \ref{sect:analytic_bondi} we qualitatively discuss the effects of a split-monopolar magnetic field on steady 
Bondi accretion. 
In Section \ref{sect:results} we describe the results obtained from our numerical simulations. In Section \ref{sect:discussions} we discuss the possible mechanism behind SASI and 
its astrophysical implications (in particular, QPOs), and summarize in Section \ref{sect:summary}.

% ................................................................................................. SASI and why ...........................................................................................................................%

\section{What is SASI and why?}
 \label{sect:mechanism}
Two different mechanisms, namely advective-acoustic and acoustic mechanisms, have been proposed to explain SASI.
Most recent studies (e.g. \citealt{Foglizzo2007}, \citealt{Foglizzo2009}) favour the former.
For a comparative and detailed discussion of the two mechanisms, see \cite{Guilet2012}. 
 \subsection {Advective-acoustic cycle}
Advective-acoustic cycle was first proposed by \cite{Foglizzo_tagger2000} in the context of Bondi-Hoyle--Lyttleton accretion. Two different waves -- an outward propagating acoustic
wave and an inward propagating entropy-vorticity wave -- contribute to this mechanism and complete a single cycle (\cite{Foglizzo2002}, \cite{Foglizzo2007}). Due to the compression of gas 
in the post-shock region (specially near the surface of neutron star), an acoustic wave is produced. The acoustic wave (propagation direction need not be purely radial) reaching the shock surface distorts it. The distortion of the shock surface, in turn, creates entropy-vorticity wave which advects down to the central neutron star and decelerates near the surface. 
Deceleration creates a positive acoustic feedback which completes the cycle. Over many cycles, the instability attains an exponential growth. With appropriate boundary conditions 
(like ours in this paper) the system reaches a quasi-steady state with stable non-linear oscillations.
 
\subsection{Acoustic cycle}
Acoustic cycle is thought to be driven by a trapped acoustic wave in the post-shock cavity. \citet{Blondin_mezzacappa2006} proposed that any density inhomogeneity  
produces sound waves near the shock surface. Due to refraction, these sound waves propagate around the circumference  of the shock until they meet on the other side. There
 their excess pressure produces a shock deformation which sends another pair of sound waves back again. The growth of the mode depends on how  pressure perturbation in the 
post shock region interacts  with the shock front.

A comparison of sonic and advection time scales should help us to distinguish these two mechanisms.

%.......................................................................................................................Method.....................................................................................................................%

\section{Method}
\label{sect:method}
To study SASI, we set up an initial value problem, in which a central accretor (e. g., a neutron star) is embedded in a stationary, spherically-symmetric uniform medium. We solve 
the magneto-hydrodynamic (MHD) equations to study the problem. 
\subsection{Equations solved}
\label{sect:physcial}
We use the {\tt PLUTO} code (\citealt{Mignone2007}) to solve the Newtonian MHD equations in spherical
coordinates ($r,\theta,\phi$). The equations are 

 \bea 
 \label{eq:mass}
&& \frac{\partial \rho}{\partial t} + \nabla .(\rho \textbf{v})= 0, \\
\label{eq:momentum}
&& \frac{\partial }{\partial t}\left(\rho \textbf{v}\right)+ 
\nabla . \left( \rho \textbf{v}\textbf{v} - \textbf{B}\textbf{B} \right)=  -\rho \nabla \Phi - \nabla P^*, \\
%\nonumber
\label{eq:energy}
&& \frac{\partial E}{\partial t} + \nabla . \left((E + P^*)\textbf{v} - \textbf{B}(\textbf{B}\textbf{v})   \right)= -\rho \nabla \Phi.\textbf{v}, \\
%\nonumber
\label{eq:induction}
&& \frac{\partial \textbf{B}}{\partial t} + \nabla . \left (\textbf{v} \textbf{B} - \textbf{B}\textbf{v} \right) = 0,
\eea
where $\rho$ is the gas density, $\textbf{v}$ is the velocity, $\textbf{B}$ is the magnetic field (a factor of $1/\sqrt{4 \pi}$ is absorbed in the definition of  $\textbf{B}$),  
$P^* = P+B^2/2$ is the total pressure ($P$ is gas pressure), and $E$ is the total energy density related to the internal energy density $\epsilon$ as $E = \epsilon + \rho v^2/2 + B^2/2$.
The adiabatic index $\gamma$ relating pressure and internal energy density ($P=[\gamma-1]\epsilon$) is chosen to be 1.4. 
Gravitational potential due to the central accretor is given by the Newtonian potential due to a point mass at the origin, $\Phi = -GM/r$. 

{\tt PLUTO} uses a Godunov type scheme  which solves the equations in conservative form. We use  the HLLD solver with second-order slope limited reconstruction. For 
time-integration, second order Runge-Kutta (RK2) is used with a CFL number of $0.4$. Divergence free constraint on magnetic field is enforced by solving a modified system of 
conservation laws, in which the induction equation is coupled to a generalized Lagrange multiplier (GLM; \citealt{Dedner2002,Mignone2010}). In this scheme, magnetic fields retain a cell centered representation.

We solve Eqs. (\ref{eq:mass})--(\ref{eq:induction}) in dimensionless form;
we express our results in both code units (e.g., timescales are expressed in units of $r_g/c$, 
$r_g = GM/c^2$) and in CGS units. In the latter case we use 
the central compact object mass to be $1M_\odot$. It is straightforward to convert from one system of units to another.

\subsection{Grid and boundary conditions}
Our spherical computational domain ($r,\theta,\phi$) extends from an inner boundary $r_{\rm in} = 6r_g$ to an outer boundary $r_{\rm out}=10^4 r_g$ in the radial direction and from $0$ to $\pi$ in the
meridional ($\theta$) direction. Here, $r_g=GM/c^2$ is the gravitational radius, where $G$ is gravitational constant and $M$ is mass of the central accretor. We use two logarithmic grids 
along radial direction, one from $r_{\rm in }$ to  $50 r_g$ with 512 grid points and another from $50r_g$ to $r_{\rm out}$ with 256 grid points. In the meridional direction 
we use a uniform grid with 256 grid points.

We fix the values of velocity components at the inner boundary; radial component $v_{r}$ is set to $v_{\rm in}$, whereas meridional component $v_{\theta}=0$ (we obtain 
similar results even if $v_\theta$ is copied in the inner radial ghost zones). 
The fiducial value of $v_{\rm in}$ is $0.05c$, but we change it to control the equilibrium shock radius. Density, pressure and magnetic field components in the ghost zones
are copied from the last computational zone near the inner 
boundary. At the outer boundary, the values of pressure, density and velocity field components are set to their initial values. The values of magnetic field components in the outer ghost 
zones are copied from the last computational zone. Axisymmetric boundary conditions (scalars and tangential components of vector fields are copied and normal components of 
vector fields are reflected) are used at both the $\theta$ boundaries ($\theta=0,\pi$).

\subsection{Initial conditions}
\label{sect:init}
We carry out 2D, axisymmetric MHD simulations in spherical ($r,\theta,\phi$) co-ordinates in an initially static ($v_r=v_{\theta}=0$) uniform ambient medium of density 
$\rho_{\rm ini}$. Initial pressure of the medium is also uniform and is given by $p_{\rm ini} = \rho_{\rm ini} c^2_{s \infty}$.  We choose the value of $c^2_{s \infty}$ 
to be $0.002 \gamma c^2$ to mimic the typical proton temperature ($\gtrsim 10^{11}$K) of the sub-Keplerian hot flow in X-ray binaries (XRBs; \citealt{Narayan_Yi1995}, \citealt{Rajesh2010}). Moreover, this choice of temperature
gives rise to a sonic radius $r_{c} \approx 71.43 r_g$ and the Bondi radius $r_{B}\approx 714 r_g$, which are well inside the computational domain. 
We initialize a split monopolar magnetic field given by,
\be
 \label{eq:mag_field}
  B_{r} = \frac{C}{r^2} {\rm sign} (\cos \theta)
\ee
The advantage of using this magnetic field configuration is that the flow structure is expected to change only close to the central accretor (i. e., at small $r$  where the field 
is strong), whereas at larger radii the solution remains unaffected (see Section \ref{sect:analytic_bondi}). The strength of magnetic field is determined by the value of the constant $C$.

\begin{table*}
\centering
\begin{tabular}{ |p{1.cm}||p{1.9cm}||p{1.2cm}||p{1.2cm}|p{1.2cm}||p{1.2cm}|p{1.2cm}|p{1.2cm}|p{1.2cm}|p{1.2cm}|p{1.2cm}|  }
 \hline
% \multicolumn{8}{|c|}\\
 \multicolumn{11}{|c|}{Simulation details} \\
 \hline
 $v_{\rm in}$     & $C$ & Label & $r_{\rm sh}/r_{\rm in}$ & $T_{a1}$ & $T_{v_{\theta}}$ & $t_{\rm aac}$ & $t_{\rm aacA+}$    & $t_{\rm aacA-}$ & $t_{\rm rcs}$ & $t_{\rm mcs}$      \\
$(c)$     &    &      &         & $(r_g/c)$   &  $(r_g/c)$    & $(r_g/c)$   & $(r_g/c)$     & $(r_g/c)$    & $(r_g/c)$     & $(r_g/c)$    \\
 \hline
 0.045 	& 1      & HD    & 4.65    & 808.17    &  808.90     & 732.55   & --  & --  & 472.73  & 718.07      \\
 0.048   	& 1      & HD    & 4.17    & 656.19    &  658.56     & 621.11   & --  & --  & 400.62  & 619.26      \\
 0.05$^\dag$   	& 1      & HD    & 3.91    & 580.24    &  581.62     & 567.42   & --  & --  & 368.08  & 567.81      \\
 0.06          & 1      & HD    & 2.96    & 351.65    &  351.79     & 359.07   & --  & --  & 234.01  & 386.90       \\
 0.07   	& 1      & HD    & 2.35    & 240.48    &  240.46     & 242.84   & --  & --  & 159.39  & 280.29       \\
\hline
 0.05          & $10^6$               & MHD  & 3.90    & 579.64     & 580.35    & 567.83   & 567.17   & 568.50   & 368.70   & 566.65 \\ 
 0.05  	& $1.25 \times 10^7$   & MHD  & 3.90    & 581.97     & 581.89    & 563.08   & 555.20   & 572.57   & 364.20   & 566.28 \\
 0.05  	& $1.67 \times 10^7$   & MHD  & 3.91    & 593.26     & 592.93    & 569.09   & 555.99   & 593.25   & 369.25   & 565.96 \\
 0.05  	& $2 \times 10^7$      & MHD  & 3.91    & 595.24     & 595.73    & 554.65   & 539.57   & 582.20   & 354.55   & 566.33 \\
 0.05  	& $2.5 \times 10^7$    & MHD  & 3.90   & 600.91     & 600.66    & 561.84   & 543.36   & 595.85   & 362.34   & 562.42 \\
 0.05  	& $3.33 \times 10^7$   & MHD  & 3.92   & 598.06     & 599.36    & 555.85   & 534.58   & 597.72   & 356.19   & 567.68 \\
 0.05  	& $4 \times 10^7$      & MHD  & 3.93 & 601.56     & 601.03    & 555.91   & 531.55   & 604.24   & 355.72   & 570.24 \\
 0.05$^\ddag$  	& $5 \times 10^7$      & MHD  & 3.94    & 617.01     & 615.07    & 565.36   & 534.40   & 630.84   & 363.61   & 571.29 \\
 0.05  	& $5.55 \times 10^7$   & MHD  & 3.95    & 630.83     & 629.70    & 575.86   & 541.71   & 642.61   & 372.54   & 572.71 \\
 0.05  	& $6.68 \times 10^7$   & MHD  & 3.96    & 651.40     & 652.32    & 580.70   & 539.97   & 659.55   & 375.80   & 573.30 \\
 0.05  	& $7.69 \times 10^7$   & MHD  & 3.98    & 667.81     & 666.94    & 583.06   & 536.02   & 680.49   & 376.37   & 576.00 \\
 0.05  	& $7.94 \times 10^7$   & MHD  & 3.98    & 672.25     & 671.98    & 584.13   & 536.94   & 694.75   & 376.87   & 576.96 \\
 \hline
\hline
 %\label{tab:simtab}
\end{tabular}
$^\dag$ The fiducial hydro run.\\
$^\ddag$ The fiducial MHD run (Case I in Section \ref{sect:flow_evolution}).\\
$^{**}$ Two MHD runs used only for QPO analysis with $v_{\rm in}=0.06c$ and  $v_{\rm in}=0.048c$ respectively, are not listed in the table.
\caption{The radial velocity at the inner boundary, $v_{\rm in}$, mainly determines the mean shock radius $r_{\rm sh}$;
$r_{\rm sh}$ is calculated by taking the time average of $a_0$ (see Eq. (\ref{eq:al})) in quasi-steady state.
The inner boundary is fixed at $r_{\rm in} = 6$ $r_g$; $C$ determines the magnetic field strength; very small value of $C$ implies 
that we are in the HD limit; SASI time period is measured by two different methods (see Section \ref{sect:sasi_period_measure}). 
Timescales related to the advective-acoustic mechanism are: $t_{\rm aac}$, $t_{\rm aacA+}$, 
$t_{\rm aacA-}$ (Eqs. (\ref{eq:taac}), (\ref{eq:taac_A_plus}), (\ref{eq:taac_A_minus})) and 
that related to the purely acoustic mechanism are: $t_{\rm rcs}$, $t_{\rm mcs}$ (Eqs. (\ref{eq:tcs}), (\ref{eq:tmcs})). }
\label{tab:simtab}
\end{table*}

%................................................................................................................Analytical Bondi.................................................................................................................% 
\section{Bondi accretion with split-monopolar field}
\label{sect:analytic_bondi}
Before discussing the simulation results, we want to investigate the effects of magnetic field configuration given in Eq. (\ref{eq:mag_field}) on the standard Bondi accretion. 
Taking the spherically symmetric form of Eqs. (\ref{eq:mass}) and (\ref{eq:momentum}) and using a polytropic equation of state, 
$P = K \rho^{\gamma}$ ($K$ is a constant related to entropy), and rearranging, we get the following set of equations
\bea
\label{eq:bondi_eq}
 &&  \frac{dv}{dr} = \frac{ \frac{2c^2_s}{r}  - \frac{GM}{r^2}  }{v - \frac{c^2_s}{v}},\\
&& \frac{dc_s}{dr} = -\frac{c_s}{2n} \left[\frac{1}{v}\frac{dv}{dr} +\frac{2}{r} \right],
\eea
where $n = 1/(\gamma -1)$ and $c_s(r)$ is the adiabatic sound speed given by $c_s = \sqrt{\gamma P/\rho}$.
\\
Eq. (\ref{eq:bondi_eq}) has a critical point (sonic point) where $c_s = v$. The location of the sonic point $r_c$ can be obtained if we set the numerator of  
Eq. (\ref{eq:bondi_eq}) to zero to avoid divergence, namely

\be
\label{eq:sonic_point}
r_c = \frac{G M}{2c^2_{sc}},
\ee
where $c_{sc}$ is sound speed at the critical point. Note that
the expression for $r_c$ is identical to the hydrodynamic Bondi solution. So the presence of a split-monopolar magnetic field does not affect the steady 
spherically symmetric accretion solution. Physically, the current is concentrated in the equator where the field vanishes and therefore ${\bf J} \times {\bf B}$ force
vanishes everywhere. But if spherical symmetry is broken, as it happens due to SASI, magnetic fields will have an effect especially at smaller radii where the
field strength is large.

%..................................................................................................................Results ...................................................................................................................%
\section{Results}
\label{sect:results}

In this section we present results from our simulations with and without magnetic fields. We begin with results in the hydrodynamic limit.

\subsection{HD}
To study SASI in the HD regime, we choose $C$ in Eq. (\ref{eq:mag_field}) to be very small 
such that the terms involving magnetic field in Eqs. (\ref{eq:momentum}) and (\ref{eq:energy}) vanish. We run simulations to study unmagnetized
SASI with five different radial velocities imposed at the inner radial ghost zones ($v_{\rm in}$; see Table \ref{tab:simtab}). 
We change $v_{\rm in}$ to control the mean shock radius $r_{\rm sh}$, a larger value of $v_{\rm in}$ gives rise to smaller $r_{\rm sh}$ 
(for details see Section 5.1 of Paper I).  This way we can 
study SASI for different values of $r_{\rm sh}/r_{\rm in}$.

\subsubsection{Flow evolution}
\label{sect:hd_evolution}

\begin{figure*}
   \includegraphics[scale=0.69]{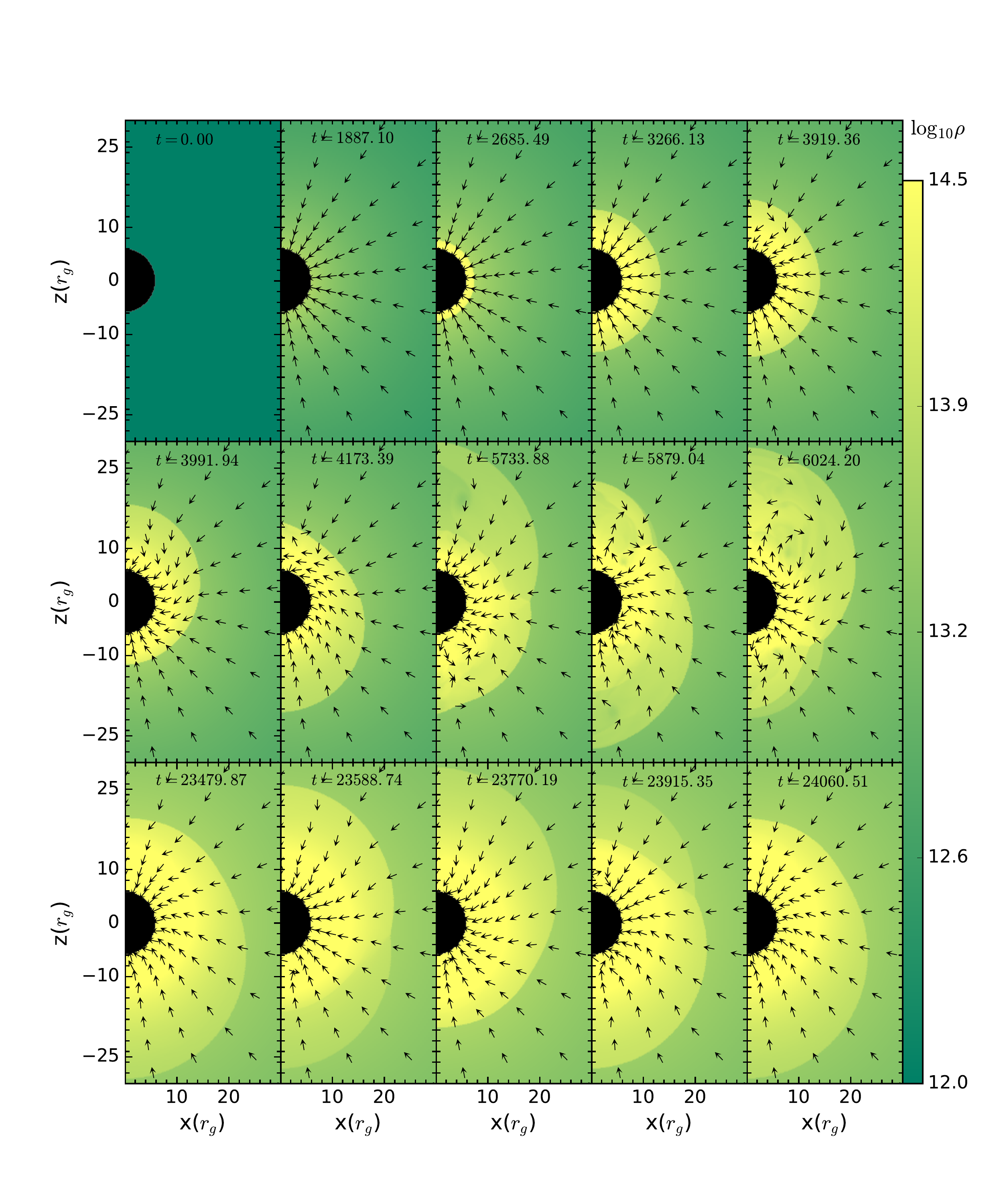}
    \caption{Density snapshots of the unmagnetized fiducial run at different time. Arrows represent the direction of velocity. Top-left panel shows the 
     initial uniform density distribution. Due to accretion of matter from the surrounding medium, both density and pressure increase, which is shown in the 
    second panel. Unlike the moderately magnetized case (c.f. Fig. \ref{fig:be2e4_rho}), the shock appears at a later time. 
    The snapshot at $t=2685.49$ ($t$ is in units of $r_g/c$) shows the first development of the shock 
    surface. The next four panels (from $t=3266.13$ to $t=4173.39$) show radial expansion of the shock surface as well as the initial build up of the vertical 
    oscillation modes. Then the post-shock cavity goes through a vigorously oscillating phase (snapshots at $t=5733.88$, $t=5879.04$, $t=6024.20$). 
    Finally, the system enters a quasi-stationary nonlinear phase in which the  post-shock cavity oscillates with a definite time period. The last five panels show a 
    full period of coherent oscillations of global modes.}
     \label{fig:be1e12_rho}
   \end{figure*}

FIG.\ref{fig:be1e12_rho} shows the density snapshots at different times for our fiducial run of unmagnetized SASI.  The details of flow 
evolution in an unmagnetized medium are described in Paper I, here we only give a brief description. We can divide the time evolution into three phases: the early 
non-equilibrium phase, 
the intermediate transition phase and the final quasi-stationary oscillating nonlinear phase. 
At $t=0$, the ambient medium is uniform and static. As the central gravitating object starts accreting, 
matter attains supersonic velocity. Both density and pressure build up near the accretor. 
Unlike classical Bondi accretion, here the supersonic matter falling under gravity 
feels an obstruction at the  inner boundary as the radial velocity there is fixed at $v_{\rm in}$. 

The accretion shock can be easily seen at $t=2685.49$ $r_g/c$. 
With time, thermal pressure builds up behind the shock due to the conversion of kinetic energy to thermal 
energy and shock surface starts expanding. The initial expansion is purely radial, but with time the radial expansion is accompanied 
by  non-spherical global oscillations with $l=1$ and higher order modes. This can be seen in the snapshots at $t=3919.36$ $r_g/c$, $t=3991.94$ $r_g/c$, $t=4173.39$ $r_g/c$.  
As the shock becomes aspherical, it becomes oblique, resulting in the generation of meridional component of velocity ($v_{\theta}$) 
in the post-shock region (see the change in direction of velocity arrows in 
the post-shock 
region for the snapshots at and after $t=3919.36$ $r_g/c$), as the mass flux ($\rho v_{\perp}$) and the tangential component of velocity ($v_{||}$) have
to be conserved across the shock. Due to the build up of  thermal pressure, the shock overcomes the inward gravitational pull and the post-shock cavity 
expands out (see snapshots at $t=5733.88$ $r_g/c$, $t=5879.04$ $r_g/c$, $t=6024.20$ $r_g/c$.). With the advection of mass and thermal energy  
across the inner boundary,  after  a few adjustments the systems attains a quasi-stationary state, in which the inward gravitational pull is balanced by outward thermal pressure.
In this state the post-shock cavity incessantly oscillates about the equatorial plane (the last five panels in FIG. \ref{fig:be1e12_rho} show one full oscillation period).

The equilibrium standing shock is linearly unstable to aspherical SASI modes but nonlinearly the systems settles into stable, long-lived, large-amplitude oscillations. The 
effective potential for such oscillations can be thought of as a local maximum within a stable potential well experienced at large amplitudes.

\subsubsection{Mode analysis}
\label{sect:mode_analysis}

\begin{figure}
    %\centering
    \includegraphics[scale=0.4]{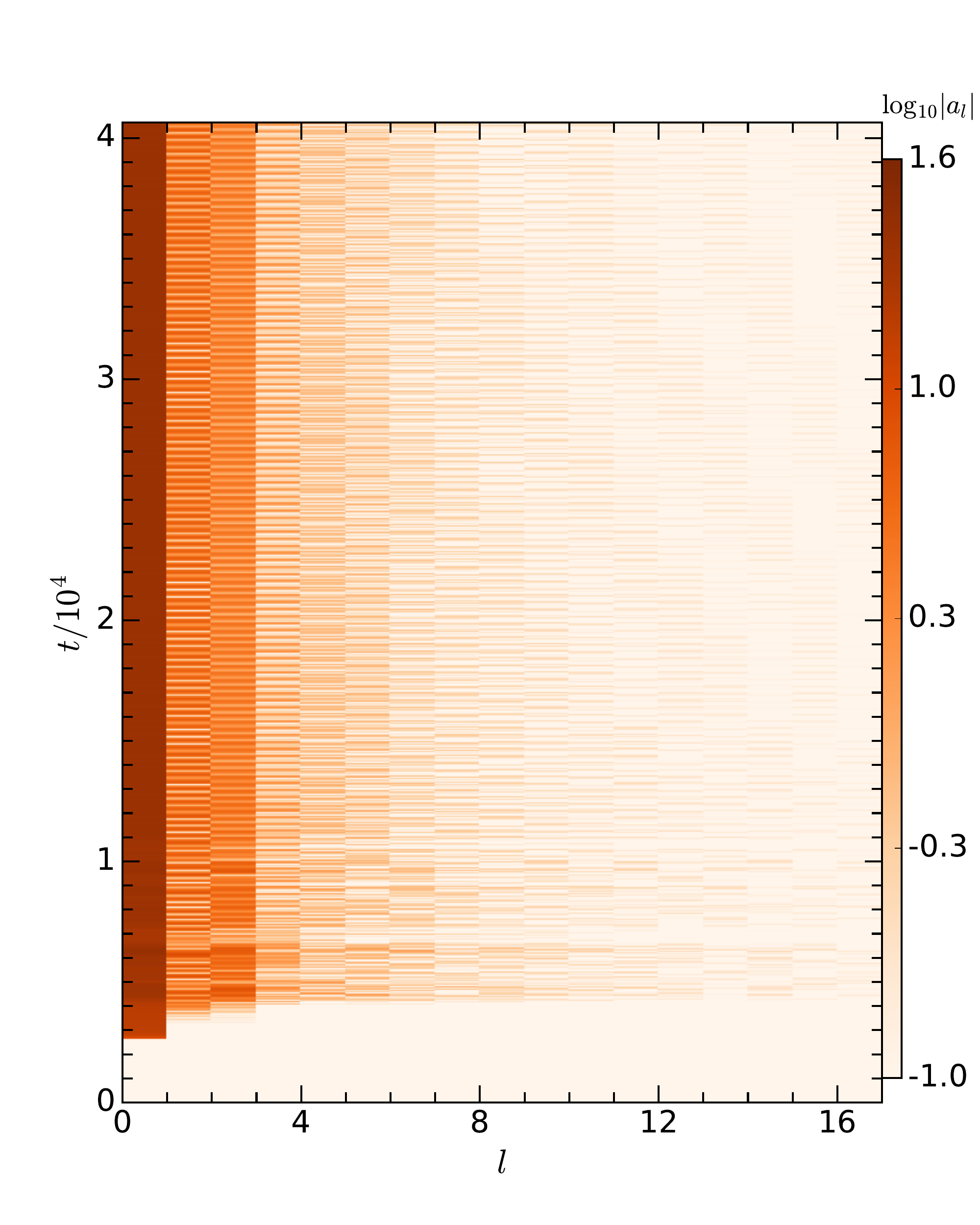}
    \caption{Result of mode analysis of the shock radius for our fiducial unmagnetized SASI run. Colour represents the absolute 
value of mode amplitude. In the quasi-stationary state, apart from the dominant $l=0$ mode, $l=1$, $l=2$  and $l=3$ modes are prominent.}
   \label{fig:mode_c_1}
 \end{figure}
    
Shock surface can be easily identified just by looking at the density jumps  in different snapshots of FIG. \ref{fig:be1e12_rho}. We see that in the nonlinear 
quasi-stationary state, shock surface can be considered as a sphere with sub-structure on top of it. To quantify the sub-structures, 
we perform mode analysis using the method of {\em spherical harmonics decomposition} (SHD).

Any spherical function $f(\theta, \phi)$ can be expanded as a linear combination of spherical harmonics as
\be
\label{eq:sh}
f(\theta, \phi) = \sum_{l=0}^{\infty} \sum_{m=-l}^{l} a_{lm}Y_{lm}(\theta,\phi),
\ee
where the spherical harmonics $Y_{lm}(\theta,\phi)$ are given by,
\be
Y_{lm}(\theta,\phi) = \sqrt{\frac{(2l+1)(l-|m|)!}{4\pi (l+|m|)!}} P_{lm}({\rm cos}\theta) e^{im\phi},
\ee
where $P_{lm}({\rm cos}\theta)$ are the associated Legendre polynomials. For an axisymmetric system, $m=0$ and $Y_{lm}$ reduces to Legendre polynomial 
$P_{l}({\rm cos}\theta)$ with a normalization factor. Then the deformed (from spherical shape) shock surface $R_s(\theta)$ can be decomposed as
\be
R_s(\theta) = \sum_{l=0}^{\infty}a_{l}P_{l}({\rm cos}\theta),
\ee
where the coefficients can be calculated as
\be
\label{eq:al}
a_{l} = \frac{2l+1}{2} \int_{\theta=0}^{\pi}R_s(\theta) P_{l}({\rm cos}\theta) {\rm sin}\theta d\theta.
\ee

FIG. \ref{fig:mode_c_1} shows the  time evolution of mode amplitudes $a_l$ for the fiducial unmagnetized SASI run. $R_s(\theta)$ is computed using the
pressure jump across the shock. Initially, the vanishing mode amplitudes reflect the absence of a shock. The first emergence of shock is reflected in the non-vanishing value of 
$a_0$, while other mode amplitudes are still zero, as the shock is spherical. As the shock starts oscillating vertically about the equatorial plane, it becomes 
aspherical in nature and $l=1$ and $l=2$ modes become prominent. In the fully nonlinear regime, we see that apart from $l=0$ mode, $l=1$, $l=2$ and $l=3$ are the most
prominent modes present. The higher order modes (specially $l\leq8$) are also present but with a smaller amplitude.

\subsubsection{Methods to measure SASI time period}
\label{sect:sasi_period_measure}

\begin{figure}
   %\centering
    \includegraphics[scale=0.36]{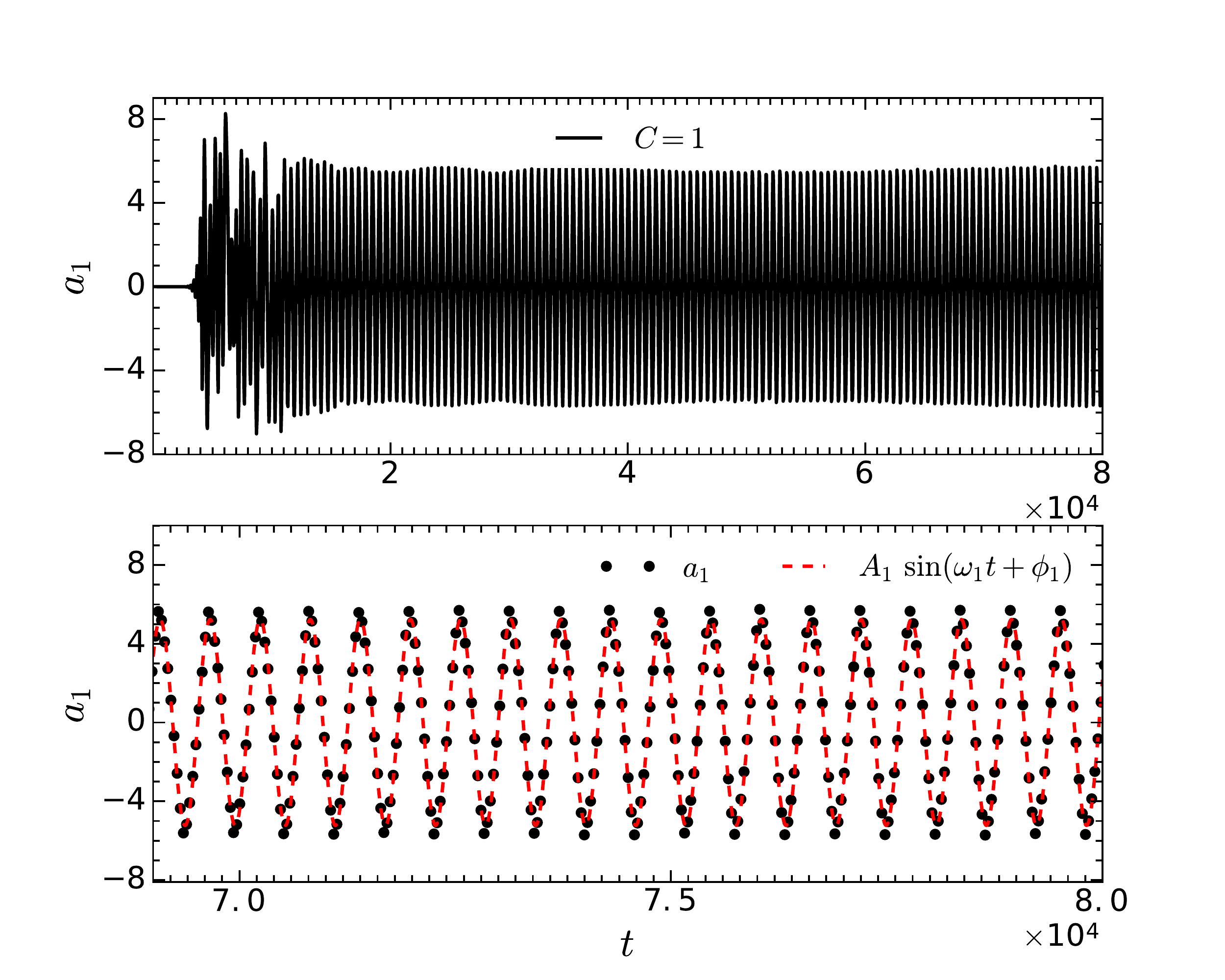}
    \caption{Top panel shows the temporal evolution of the $l=1$ component of the shock radius ($a_1$) for the fiducial unmagnetized 
SASI run. To find the time period of oscillations of global modes, $a_1(t)$ is fit with a function
$\psi_1 = A_1 {\rm sin}(\omega_1 t + \phi_1)$. The best-fit time period  is $T_{a1}= 2 \pi/\omega_1 = 580.24$ $r_g/c$. Plots of  
$a_1(t)$ and the best-fit function are shown in the bottom panel.}
    \label{fig:amp_1e12}
    \end{figure}

\begin{figure}
   %\centering
    \includegraphics[scale=0.35]{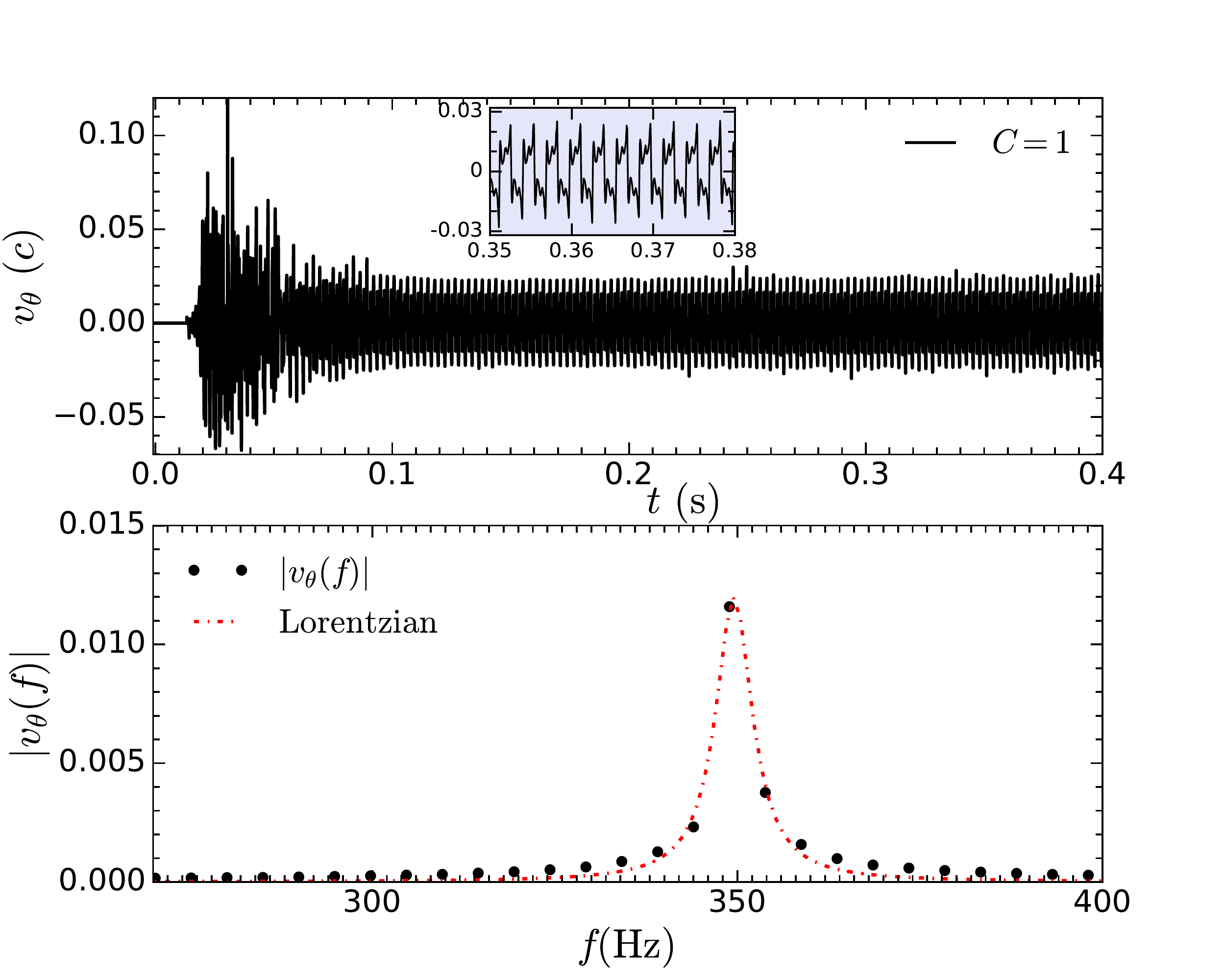}
    \caption{Top panel shows the temporal evolution of $v_{\theta}$ at a fixed location ($r=8 r_g/c,~\theta=\pi/2$) for  the fiducial HD run. 
    FFT of $v_{\theta}(t)$ (bottom panel) fitted with a Lorentzian (Eq. \ref{eq:Lorentzian}) gives the time period of $v_{\theta}(t)$ . 
    Time period of the oscillations is $T_{v_{\theta}}= 1/f_{v_{\theta}} = 581.62$ $r_g/c$.}
    \label{fig:tvth_1e12}
 \end{figure}

It is clear from FIG. \ref{fig:be1e12_rho} that there are global nonlinear oscillation modes associated with the post-shock cavity. We want to determine the time period of oscillations. 
%The easiest way is eye estimation, but it is tedious and as well as error-prone. 
We use two different methods to find the precise oscillation period:
 \\
(i) Following \cite{Ohnishi2006}, we fit the mode amplitude (in quasi-steady state) associated with $l=1$ with a {\em sine} curve given by, 
$\psi_1 = A_1~ {\rm sin} (\omega_1 t + \phi_1)$. Time period is obtained from the value of $\omega_1$ as $T_{a1}=2 \pi/\omega_1$. Top panel of 
FIG. \ref{fig:amp_1e12} shows the temporal variation of $a_1$ for our fiducial unmagnetized SASI run ($v_{\rm in}=0.05c$, $C=1$). After the initial growing 
phase, $a_1$ attains a quasi-steady state and oscillates about a mean value close to 0. In the bottom panel of FIG. \ref{fig:amp_1e12}, the simultaneous plots of $a_1$ and 
the fitting function $\psi_1$ are shown. The original data and the fitting function match well and the measured time period of SASI is $T_{a1} = 580.24$ $r_g/c$. 
 \\
 (ii) The second method to obtain the time period of oscillations is based on calculating the temporal variation of a local quantity at a single point in space.  We choose $v_{\theta}$ 
to be the local quantity because $v_{\theta}$ changes sign as the post-shock cavity goes from the upper hemisphere to the lower hemisphere. We compute 
$v_{\theta} (t)$ in the equatorial plane ($\theta= \pi /2$) at $r=8 r_g$ as a function of time, which is shown in the top panel of 
FIG. \ref{fig:tvth_1e12} for the fiducial HD run. Note that $v_{\theta}(t)$ is not a purely sinusoidal function (see the inset in top panel of Fig. \ref{fig:tvth_1e12}). 
To find the time period associated with it, we take the fast Fourier transform (FFT) and identify the most prominent peak in the power spectrum (defined here as simply the 
absolute value of the Fourier transform). 
To find the centroid frequency ($f_{v_{\theta}}$), we fit the prominent peak with a Lorentzian given by,
 
 \be
 \label{eq:Lorentzian}
 P(f) = \frac{A_L \Gamma}{(f-f_{v_{\theta}})^2 + \Gamma^2/4}
 \ee
 where $A_L$ is the normalization and $\Gamma$ is the full width at half maximum (FWHM). Inverse of $f_{v_{\theta}}$ gives the oscillation period ($T_{v_{\theta}}$) 
measured by this method. Bottom panel of FIG. \ref{fig:tvth_1e12} shows the power spectrum of $v_{\theta}(t)$; the fitting function $P(f)$ is plotted on top of it to compare the actual and 
fitted values of the power spectrum. The time period of the oscillations obtained by this  method is $T_{v_{\theta}}=581.62$ $r_g/c$. We note that both methods  (i) \& (ii) 
give almost identical results.    
 
 \subsubsection{Timescales from linear theory}
 \label{sect:time_scales_HD}
 We measure the time period of oscillations in the quasi-steady, nonlinear phase with large amplitude. Linear theories of SASI predict important timescales
 related to the propagation of various perturbations. While not strictly valid, the various signal propagation timescales are expected to provide an appropriate 
 scaling even for the nonlinear oscillations. The following arguments based on simple signal propagation timescales stem from the fact that the disturbances 
 have to reflect and travel back to the origin of waves to interfere and create a standing wave. In some mechanisms mode conversion (e.g., from acoustic to 
 vorticity/entropy modes and vice versa) is invoked at the boundaries.

 First, we define an advective-acoustic timescale (\citealt{Foglizzo2007}) as the sum of the radial advection time from shock surface to the inner boundary
 and the acoustic time to return back to the shock surface in radial approximation
 \be
 \label{eq:taac}
 t_{\rm aac} = \int_{r_{\rm in}}^{r_{\rm sh,O}} \frac{dr}{|\bar{v}_r(r)|} + \int_{r_{\rm in}}^{r_{\rm sh,I}} \frac{dr}{(\bar{c}_s(r) - |\bar{v}_r(r)|)},
 \ee
where $\bar{v}_r(r)$ is the $\theta$-averaged (throughout the paper we use an overline to represent angle average) radial velocity within the shock and 
the integrals are performed within the shock, in the sense that (the following also applies 
to the other radial timescales that follow; c.f. Eq. (\ref{eq:tcs}))
$$
\bar{v}_r(r) = \frac{\int_{0}^{\pi} H v_r(r,\theta) {\rm sin} \theta d\theta }{\int_{0}^{\pi} H {\rm sin} \theta d\theta },
$$ 
where, $H=H[r_{\rm sh}(\theta)-r]$ is the Heaviside step function whose value is zero for negative argument and one for positive argument.
Here we want to emphasize that there are two shock surfaces at certain times 
(e. g., see snapshot at $t=23915.35$ $r_g/c$ in FIG. \ref{fig:be1e12_rho}), and for calculating the advection 
time (or any time associated with signals propagating inward) we compute the time taken by the fluid element to reach inner boundary $r_{\rm in}$ from 
the maximum outer shock radius $r_{\rm sh,O}$. But for calculating the acoustic time (or any time associated with outward-propagating signals), we compute the time taken by 
the outward-propagating sound wave to reach the maximum inner shock radius $r_{\rm sh,I}$ from the inner boundary $r_{\rm in}$, as acoustic signals cannot propagate 
outside the shock at $r_{\rm sh,I}$. The timescales vary with time because of the finite amplitude of the shock oscillations but average timescales should be indicative 
of the fundamental mode.

Second, we compute the radial acoustic timescale, sum of the times taken by the sound waves to reach the shock surface from the inner boundary and back,

\be
 \label{eq:tcs}
 t_{\rm rcs} = \int_{r_{\rm in}}^{r_{\rm sh,O}} \frac{dr}{(\bar{c}_s(r) + |\bar{v}_r(r)|)} + \int_{r_{\rm in}}^{r_{\rm sh,I}} \frac{dr}{(\bar{c}_s(r) - |\bar{v}_r(r)|)}.
 \ee

 Third, we compute the meridional acoustic time (\citealt{Blondin_mezzacappa2006}), considering the propagation of sound wave along the circumference of the shock,
 \be
 \label{eq:tmcs}
 t_{\rm mcs} = \int_{0}^{\pi} \frac{r_{\rm sh}(\theta)}{c_s(r_{\rm sh},\theta)} d\theta.
 \ee

 \begin{figure}
   %\centering
    \includegraphics[scale=0.45]{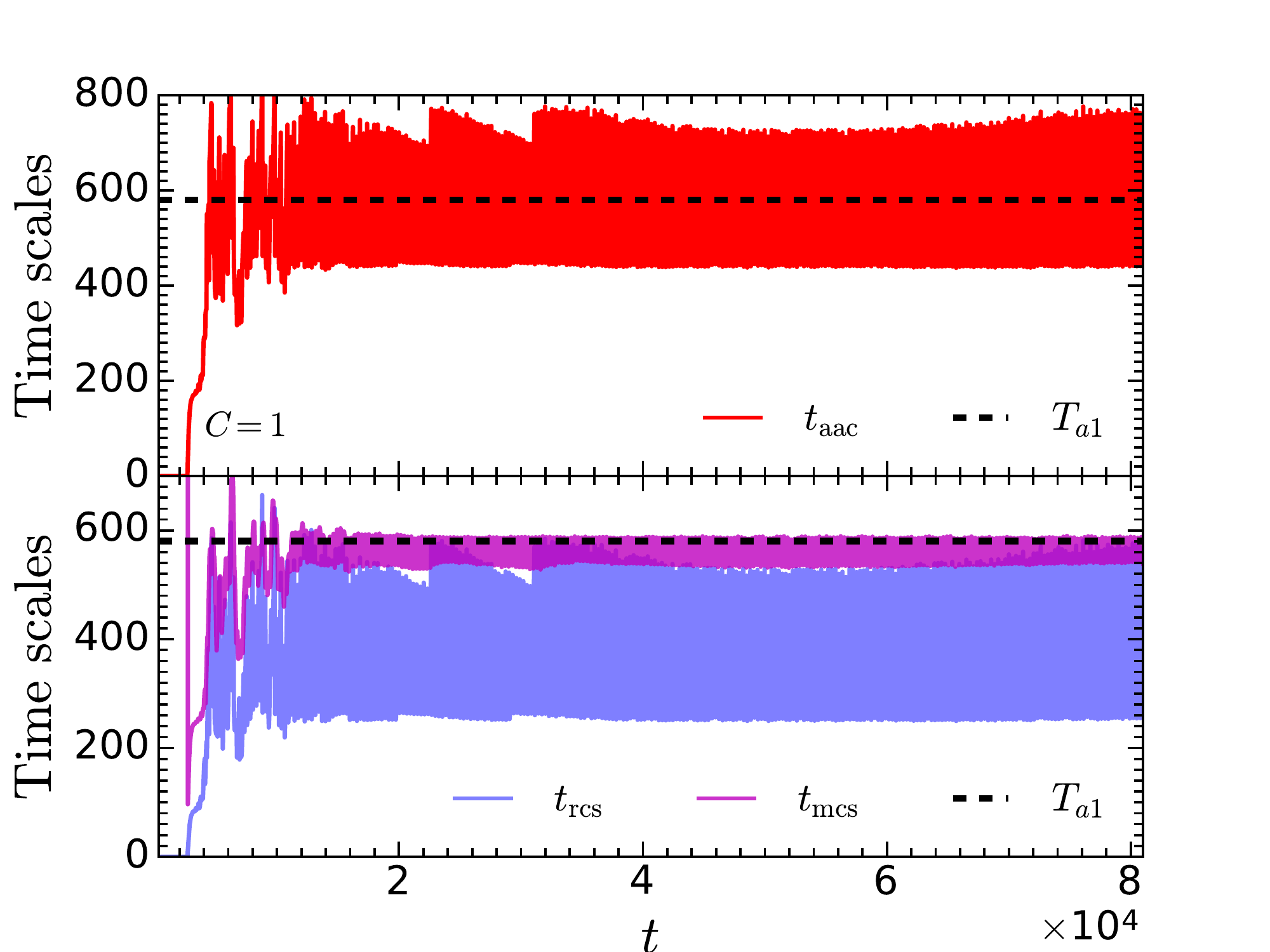}
    \caption{Comparison of different timescales obtained from linear theory (namely, advective-acoustic cycle, $t_{\rm aac}$; 
radial acoustic cycle, $t_{\rm rcs}$; meridional acoustic cycle, $t_{\rm mcs}$)  with SASI time period $T_{a1}=580.24$ for the 
fiducial model of unmagnetized SASI ($C=1$ and $v_{\rm in}=0.05c$). The time averaged (in the quasi-steady state) values of
 the timescales are $<t_{\rm aac}>=567.41$, $<t_{\rm mcs}>=567.81$, $<t_{\rm rcs}>=368.03$; time is in units of $r_g/c$.}
    \label{fig:ts_c_1}
 \end{figure}
 
 FIG. \ref{fig:ts_c_1} shows the  plots of the observed time period $T_{a1}$ (since $T_{a1}$ and $T_{v_{\theta}}$ are almost identical,
  we plot only $T_{a1}$) with the above timescales obtained from linear theory for the fiducial unmagnetized SASI simulation. 
In the quasi-steady state the theoretical timescales oscillate in time with a large amplitude
 ($~80-300$  $r_g/c$). While both the advective-acoustic time $t_{\rm aac}$ and meridional acoustic time 
$t_{\rm mcs}$ contain $T_{a1}$ within their range of variations, radial acoustic time $t_{\rm rcs}$ is shorter. 

As the variations in timescales are large, we take the time  average between $t=50000$ $r_g/c$ and $t=82000$ $r_g/c$. The time averaged values of  advective-acoustic scales 
(<$t_{\rm aac}> =567.41$ $r_g/c$) and meridional acoustic timescales (<$t_{\rm mcs}>=567.81$ $r_g/c$) are close to the observed time
 period $T_{a1}=580.24$ $r_g/c$. It is a coincidence that <$t_{\rm aac}$> and <$t_{\rm mcs}$> are so close. Note that according to 
 \cite{Blondin_mezzacappa2006} the time period of SASI oscillations is 
expected to be $2t_{\rm mcs}$ (so that the two waves originated at one point near the shock surface can interfere on the other side and return 
back to the origin), whereas  we find a close match of $t_{\rm mcs}$ to the measured SASI time period. 
On the contrary, the time averaged value of 
radial acoustic time is $<t_{\rm rcs}>=368.03$ $r_g/c$, much less than the SASI time period. So there appears to be a degeneracy between 
the two timescales $t_{\rm aac}$ and $t_{\rm mcs}$ derived from two different physical mechanisms, namely advective-acoustic and purely acoustic cycles. 

\begin{figure}
   %\centering
    \includegraphics[scale=0.45]{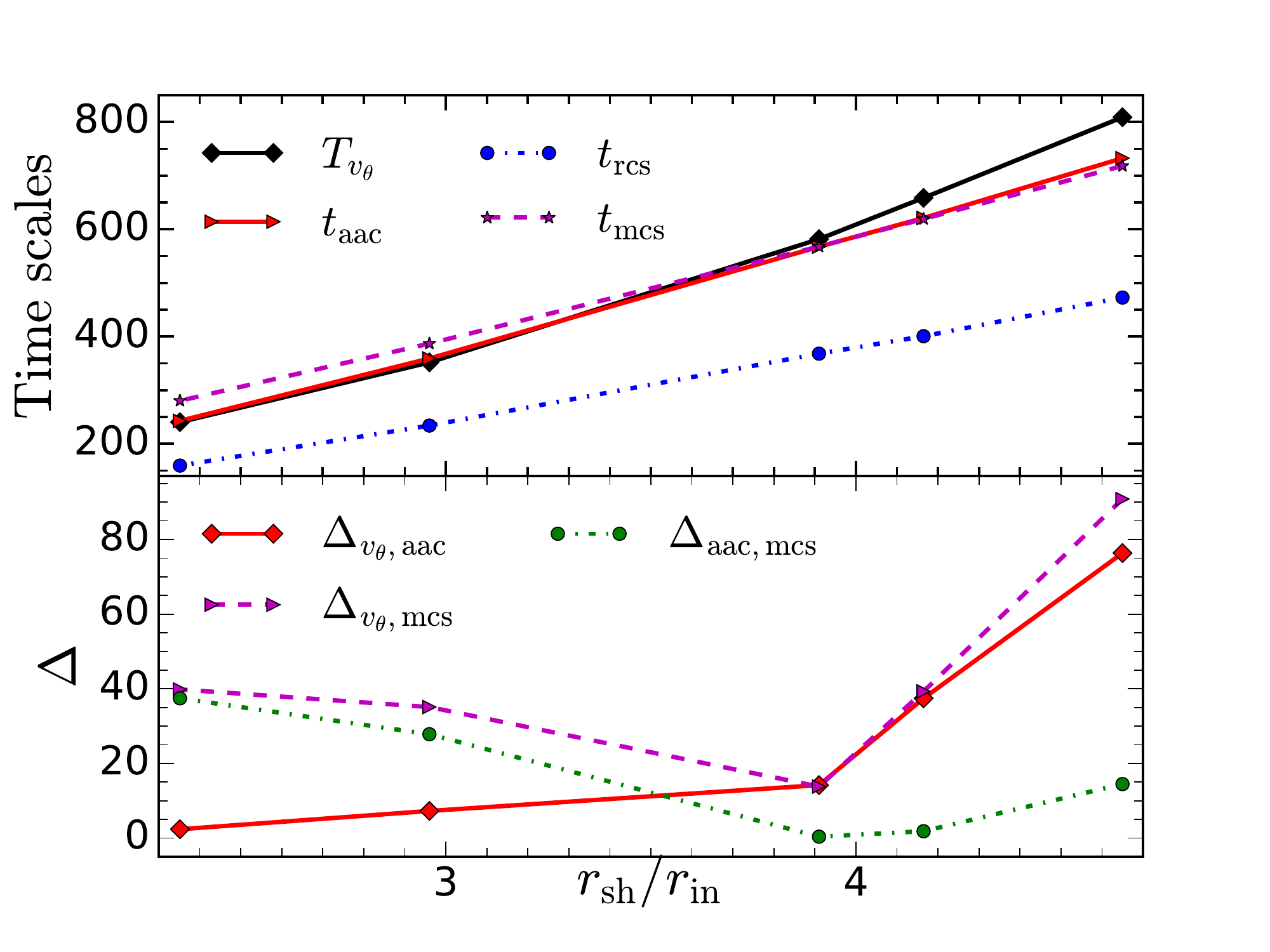}
    \caption{Top panel: Variation of different average timescales (namely, advective-acoustic cycle, $t_{\rm aac}$; radial acoustic cycle, $t_{\rm rcs}$; meridional 
     acoustic cycle, $t_{\rm mcs}$; and  SASI time period $T_{v_{\theta}}$; in units of $r_g/c$) with change in the ratio of the mean shock radius and the inner radius,
     $r_{\rm sh}/r_{\rm in}$ for unmagnetized SASI.
     Bottom panel: Change in the absolute value of difference ($\Delta$) between two timescales  with $r_{\rm sh}/r_{\rm in}$:
     $\Delta_{v_{\theta}, {\rm aac}} = |T_{v_{\theta}} - t_{\rm aac}|$; $\Delta_{v_{\theta}, {\rm mcs}} = |T_{v_{\theta}} - t_{\rm mcs}|$;
     $\Delta_{{\rm aac}, {\rm mcs}} = |t_{\rm aac} - t_{\rm mcs}|$. Note that of the various timescales, the advective-acoustic timescale most closely matches the 
     observed SASI time period.
   }
    \label{fig:rsh_rin_time}
 \end{figure}

To break this degeneracy (between $t_{\rm aac}$  and $t_{\rm mcs}$) because of our choice of parameters, we change the shock location by tuning the value of $v_{\rm in}$
 and measure  the  oscillation period as well as the relevant timescales.  
Top panel of FIG. \ref{fig:rsh_rin_time} shows the time averaged values of velocity oscillation time period ($T_{v_{\theta}}$; described in (ii) in 
Section \ref{sect:sasi_period_measure}), the advective-acoustic  time ($t_{\rm aac}$) and the meridional acoustic time ($t_{\rm mcs}$)  as a function $r_{\rm sh}/r_{\rm in}$. 
 The bottom panel of FIG. \ref{fig:rsh_rin_time} shows the absolute value of the difference between different relevant timescales --
 ($\Delta_{v_{\theta}, {\rm aac}} = |T_{v_{\theta}} - t_{\rm aac}|$; $\Delta_{v_{\theta}, {\rm mcs}} = |T_{v_{\theta}} - t_{\rm mcs}|$;
     $\Delta_{{\rm aac}, {\rm mcs}} = |t_{\rm aac} - t_{\rm mcs}|$) -- as a function of $r_{\rm sh}/r_{\rm in}$. For smaller $r_{\rm sh}/r_{\rm in}$, 
     the advective-acoustic time ($t_{\rm aac}$) matches the SASI time period measured by $T_{v_{\theta}}$. For larger  $r_{\rm sh}/r_{\rm in}$ the radial advective-acoustic 
     timescale is shorter perhaps because of non-radial propagation of sound waves. Also note the closeness
     between $t_{\rm aac}$ and $t_{\rm mcs}$ for $r_{\rm sh}/r_{\rm in} \gtrsim 3.9$, which makes it harder to choose between the two cycles in this regime.
     
\subsection{MHD}

In this section we present results from our simulations of initially split-monopolar magnetic fields with varying field strengths.

 \subsubsection{Evolution of the flow}
 \label{sect:flow_evolution}

   \begin{figure*}
 %\centering
    \includegraphics[scale=0.72]{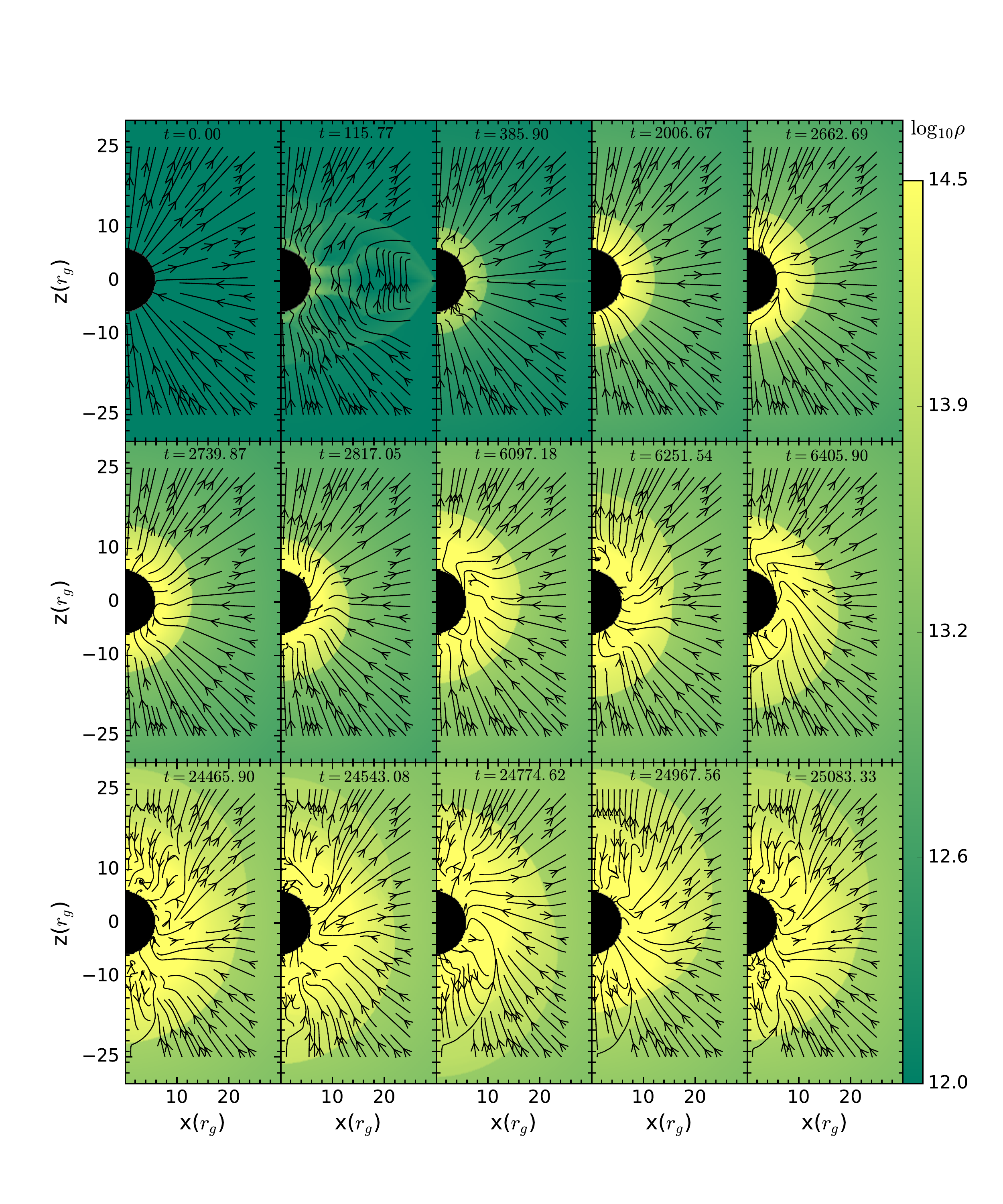}
    \caption{{\em Case I}: Evolution of the flow with time for the fiducial MHD run. Colour represents the 
density and streamlines show the magnetic field lines. First panel shows the initial uniform density distribution. A transient phase is shown in second panel. 
The next two panels  (from $t=385.90$ to $t=2006.67$) describes growing phase of post-shock cavity. A transition period from an growing phase 
to an quasi-steady phase of SASI is shown in next six panels (from $t=2662.69$ to $t=6405.90$). The last five panels shows a a full 
period of coherent oscillations of global modes. $t$ is in units of $r_g/c$.}
     \label{fig:be2e4_rho}
   \end{figure*}

  In this section we discuss the time  evolution of the magnetized flow with the radial inward velocity at the inner boundary set to $v_{\rm in}=0.05c$, 
  as in the fiducial HD run. We focus on two different cases with moderate and high field strengths.
  \\
  {\em Case I}: in which the magnetic field strength is moderate and SASI (identified by coherent shock oscillations) exists; this is the fiducial MHD run 
  with $C=5\times 10^7$ (see Eq. (\ref{eq:mag_field})) marked in Table \ref{tab:simtab}.
  \\
  {\em Case II}: in which a strong magnetic field prevents a shock from existing at late times, with $C=8\times 10^7$.
  
FIG. \ref{fig:be2e4_rho} shows the density snapshots for {\em Case I}. Streamlines 
show magnetic field lines. At $t=0$, the ambient density  is uniform and the magnetic pressure is comparable to the thermal pressure close to the accretor. 
Like the unmagnetized simulations,  the magnetized runs with moderate field strengths go through three phases: 
an early phase in which a shock develops (top panels in FIG. \ref{fig:be2e4_rho}), the intermediate transition period (middle panels in 
FIG. \ref{fig:be2e4_rho}), and a final quasi-stationary phase (bottom panels in FIG. \ref{fig:be2e4_rho}).  %In addition, 
The flow  undergoes a very early transient phase (see snapshot at $t=115.77$ $r_g/c$), %. During \sout{the linear} this phase  of evolution, 
during which thermal 
pressure builds up due to the conversion of  gravitational (via kinetic energy) to thermal energy, and the shock surface starts expanding radially outwards 
(see snapshots at $t=385.90$ $r_g/c$ and at $t=2006.67$ $r_g/c$). Finally, the shock executes coherent oscillations.

\begin{figure*}
   \centering
    \includegraphics[scale=0.9]{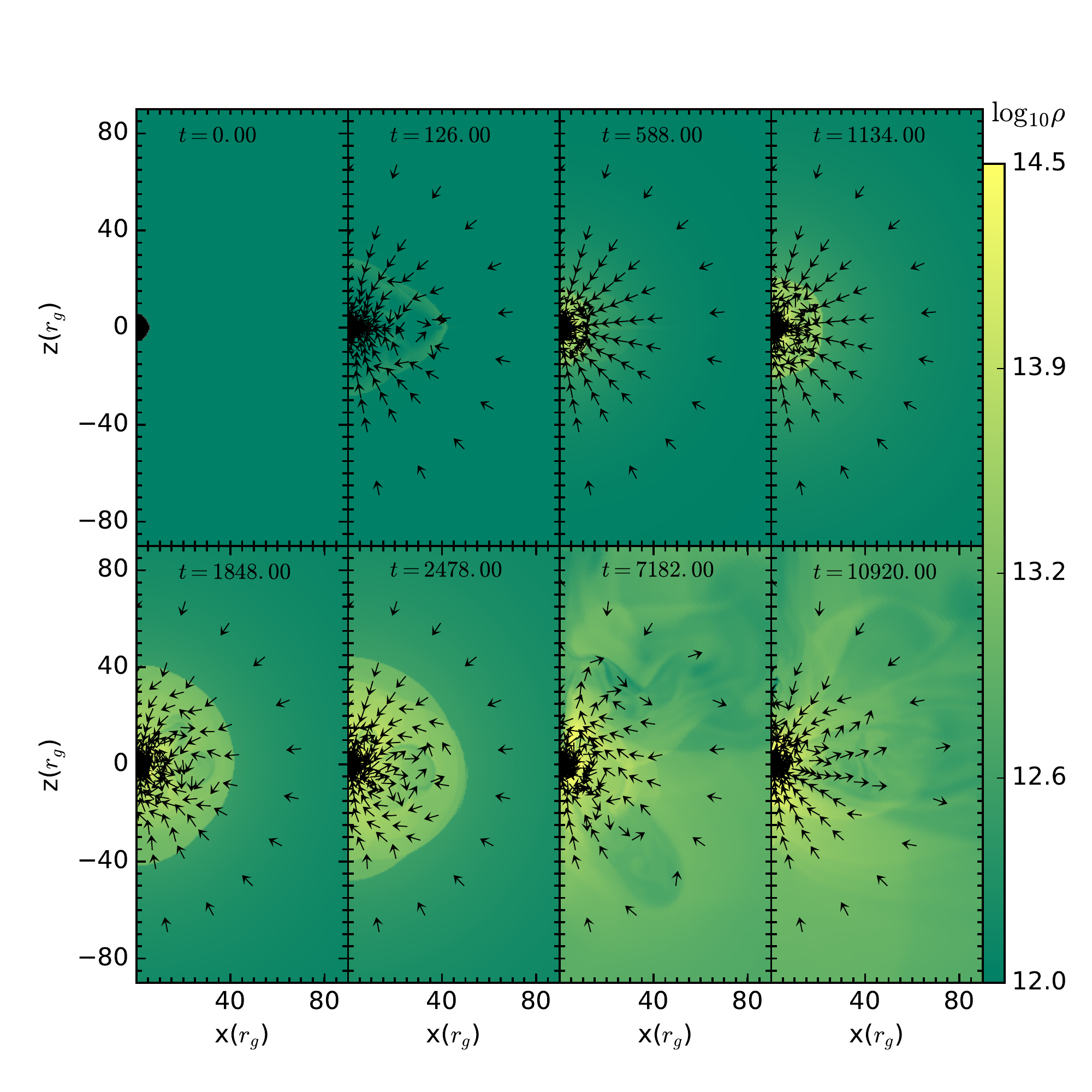}
    \caption{{\em Case II}: Evolution of the flow with time for $C= 8 \times 10^7$ and $v_{\rm in}=0.05c$. Colour represents the density and arrows 
    are for velocity directions. Starting with an initial uniform density distribution (left most top panel), system
 undergoes a transient phase (second panel on top). Then, like in {\em Case I} a spherical shock appears which grows in time ($t=588$). 
But unlike {\em Case I}, the shock does not stall, but keeps on propagating out($t=1134$ and $t=1848$) and looses its strength and vanishes ultimately 
($t=7182.00$ and $t=10920$). $t$ is in units of $r_g/c$.}
    \label{fig:be1_2e4_rho}
   \end{figure*}

Even with a slightly higher magnetic field strength ($~1.6$ times {\em Case I}), the
temporal evolution of {\em Case II} is qualitatively different because the shock is absent at late times.
FIG. \ref{fig:be1_2e4_rho} shows the density snapshots for {\em Case II}, over plotted with arrows showing the velocity unit vectors. Like 
in {\em Case I}, after undergoing a transient phase (see snapshot at $t=126$ $r_g/c$), a spherical shock is formed which expands in time (see snapshots at $t=588$ $r_g/c$ 
and $t=1134$ $r_g/c$).  However, the shock does not stall but keeps on expanding and becoming weaker, as gravitational pull is unable to balance the 
outward (thermal + magnetic) pressure. When shock reaches the sonic point ($r_c\sim 71$ $r_g/c$), the flow becomes subsonic and the shock disappears. 
Eventually, a hydrostatic atmosphere is formed. Snapshots at $t=1848$ $r_g/c$ and at 
$t=2478$ $r_g/c$ show the outward propagation of shock, while snapshots at $t=7182$ $r_g/c$ and at $t=10920$ $r_g/c$ show the flow structure 
when shock disappears. 

  \begin{figure}
   \centering
    \includegraphics[scale=0.5]{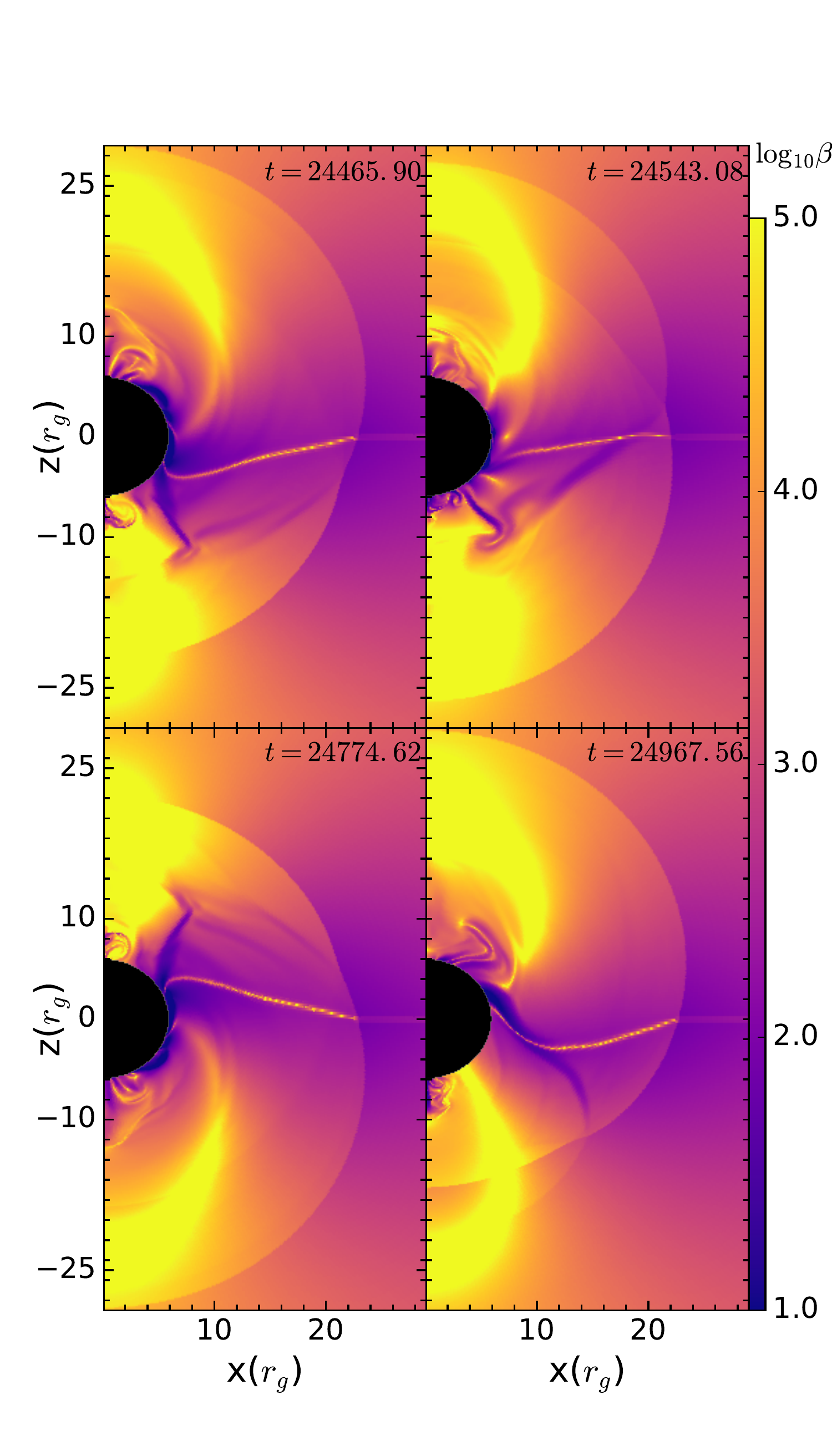}
    \caption{Snapshots of plasma $\beta~(\equiv 2P/B^2)$ at late times for {\em Case I}; $t$ is in units of $r_g/c$. 
    The color scale has been cut off at the minimum and maximum value
    shown in the colorbar. The inner and equatorial regions have a higher field strength. Close to the mid-plane, one can 
    clearly see an oscillating current sheet.}
    \label{fig:beta_snapshots}
   \end{figure}

   FIG. \ref{fig:beta_snapshots} shows the snapshots of plasma $\beta$ (the ratio of thermal and magnetic pressures) in quasi-steady state
   for {\em Case I}. Plasma $\beta$ close to the shock surface is $\gtrsim 10^3$, and therefore magnetic fields are not expected to noticeably change the shock oscillation
   period if the underlying mechanism for SASI involves meridional propagation of fast MHD waves (generalization of sound waves in the MHD regime),
   characterized by the meridional acoustic timescale $t_{\rm mcs}$ (Eq. (\ref{eq:tmcs})). Later we shall see that the SASI oscillation frequency in presence 
   of magnetic field changes noticeably (c.f. black solid line with square symbols in Fig. \ref{fig:ts_mag}), ruling out the meridional acoustic mechanism for 
   SASI.

   \begin{figure}
    %\centering
    \includegraphics[scale=0.37]{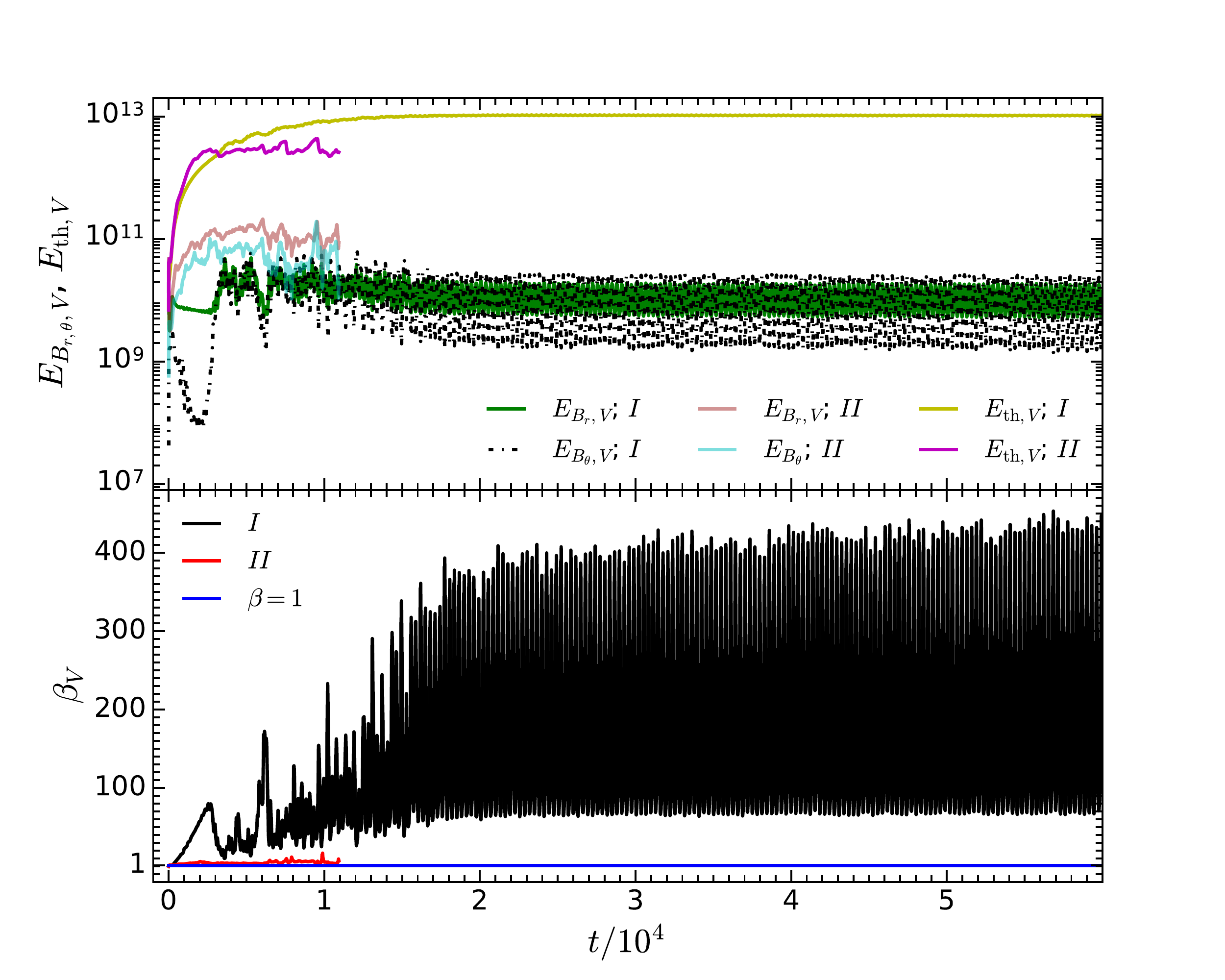}
 \caption{Top panel: Temporal evolution of average (over  the volume $V$ enclosed within $r=30 r_g$) thermal energy $E_{{\rm th},V}$ 
    and magnetic energies associated with radial ($E_{B_{r},V}$) and meridional ($E_{B_{\theta},V}$) components
   of magnetic field for {\em Case I} and {\em Case II}. Bottom panel: Temporal evolution of the $\beta_V$, the ratio of volume averaged gas  
   pressure ($P$) to magnetic pressure ($B^2/2$) for the same runs. While, time averaged $\beta_V$ in quasi-stationary state is $\sim 300$ 
   and  shock persists; in {\em Case II}, due to presence of stronger initial magnetic field, the flow gets  choked and the final  $\beta_V \sim 5$ and 
   shock disappears at late times.}
    \label{fig:beta_1}
    \end{figure}
  To quantify the magnetic field strength within the shock, we define a volume averaged quantity,
  \be
  \label{eq:beta_V}
   \beta_V = \frac{\int_V PdV}{\int_V (B^2/2)dV} = \frac{ (\gamma -1) E_{{\rm th},V}}{E_{{B_r,V}}+E_{B_{\theta},V}},
  \ee
  where the volume $V$ over which the integral is done extends from inner boundary to $r=30 r_g$; this radius is well inside the sonic 
radius $r_c$ ($=71 r_g$ for our parameters; Eq. (\ref{eq:sonic_point})), and the shock radius $r_{\rm sh}$ is always within it. Similarly, $E_{{\rm th},V}$ is the volume averaged 
thermal energy, and $E_{B_r,V}$ and $E_{B_{\theta},V}$ are  the volume averaged magnetic energies of the radial and meridional components of 
the magnetic field.
  
The top panel of FIG. \ref{fig:beta_1} shows the evolution of volume averaged magnetic 
and thermal energies within $30 r_g$ (top panel) with time  for {\em Cases I \& II}; bottom panel shows the evolution of $\beta_V$ (Eq. (\ref{eq:beta_V})). 
In the top panel of FIG. \ref{fig:beta_1}  for {\em Case I}, during purely radial expansion of post-shock cavity, thermal energy (yellow line) increases
rapidly with time due to shock heating, but magnetic energy remains roughly constant because radial flows cannot amplify a radial field. 
As a result, $\beta_{V}$ increases during this phase of evolution. Later, the radial expansion of the shock is accompanied by global oscillations with $l=1$ and higher order modes
(see snapshots at $t=2662.69$ $r_g/c$, $t=2739.87$ $r_g/c$, $t=2817.05$ $r_g/c$ in FIG. \ref{fig:be2e4_rho}). The turbulent (in transition phase) and oscillatory 
meridional velocity associated with non-spherical modes amplifies magnetic fields at later times.  
Simultaneous increase of thermal and magnetic pressure
causes further expansion of the post-shock cavity (in FIG. \ref{fig:be2e4_rho}, see snapshots at $t=6097.18$ $r_g/c$, $t=6251.54$ $r_g/c$, $t=6405.90$ $r_g/c$). 
Though, both thermal and magnetic energies increase simultaneously, the build up of magnetic energy is more erratic.
Eventually, {\em Case I} attains a quasi-stationary state, in which both thermal and magnetic energies start oscillating about a mean value. 

For {\em Case II}, magnetic field amplification happens earlier compared to {\em Case I} due to presence of aspherical shock
from the very beginning of the flow evolution (see snapshots at $t=126$ $r_g/c$, $t=588$ $r_g/c$, and at $t=1134$ $r_g/c$ in FIG. \ref{fig:be1_2e4_rho}).
Once shock disappears, magnetic energies (both $E_{B_r, V}$ and $E_{B_{\theta},V}$) saturate. This early amplification of magnetic field results in a low
$\beta_V$ (close to $\sim 5$) during the flow evolution which in turn chokes the flow. The temporal evolution of the flow in {\em Case II} is equivalent to 
a hydro set-up with reflective inner radial boundary condition, or more precisely, if $v_{\rm in}$ is smaller than the lower limit of velocity (at the inner boundary) for which a 
stationary shock solution is possible (see Fig. 15 and Section 5.1 in Paper I).

 \subsubsection{Mode analysis}
 \label{sect:mode_analysis}

   \begin{figure}
      %\centering
    \includegraphics[scale=0.4]{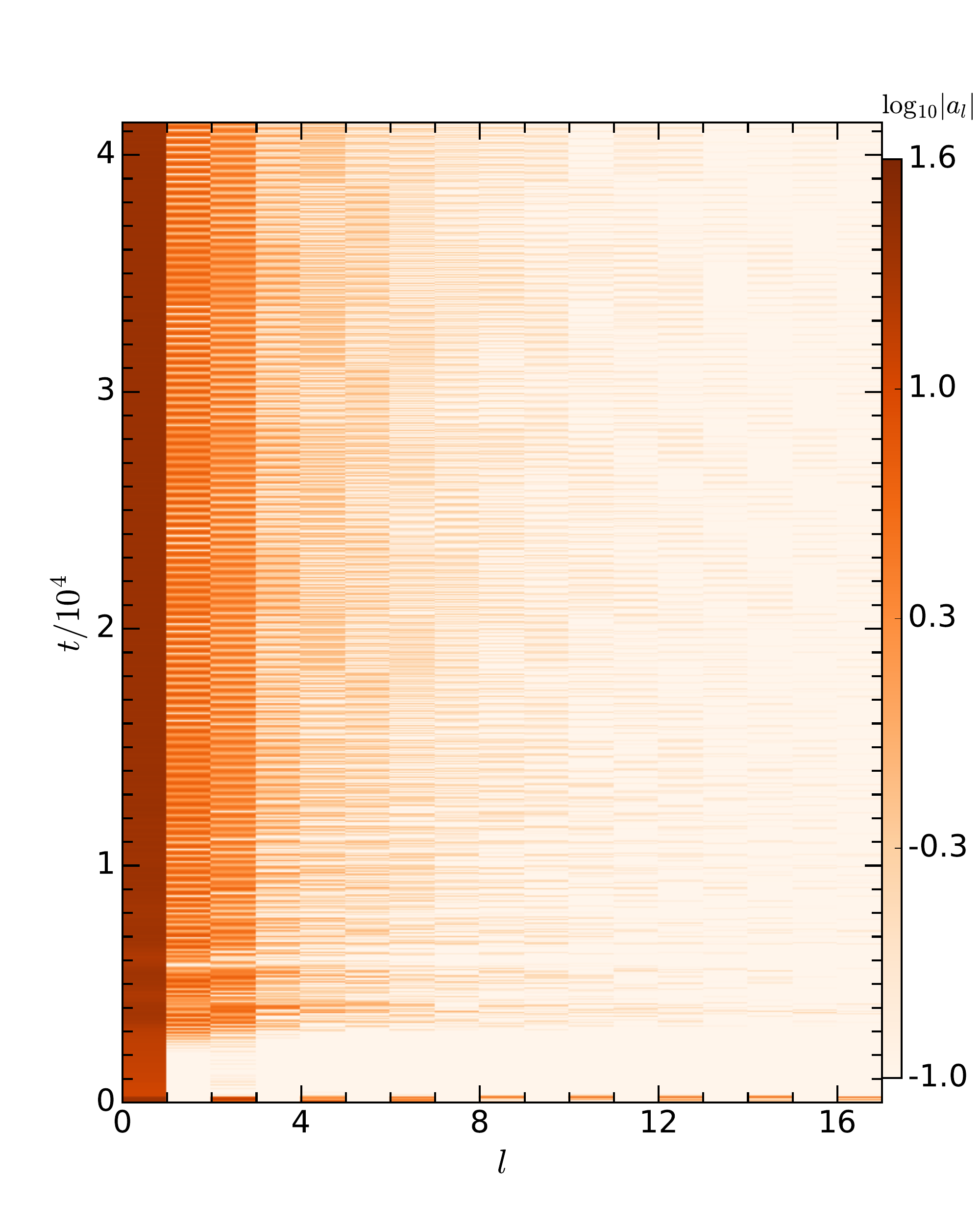}
    \caption{Results of mode analysis for the fiducial MHD run. Colour represents the absolute value of mode amplitude. 
$l=0$ is the most dominant mode at all time. During the very early transient phase, only even 
modes ($l=0,2,4,6$ etc) dominate. After the transient phase, during pure radial expansion of shock, only $l=0$ mode dominates.
As soon as shock-structure starts oscillating in the vertical direction, apart from $l=0$ mode, $l=1$, 
$l=2$ and $l=3$ modes  dominate over the higher order modes.}
   \label{fig:mode_c_5e7}
    \end{figure}
           
    \begin{figure}
     %\centering
    \includegraphics[scale=0.45]{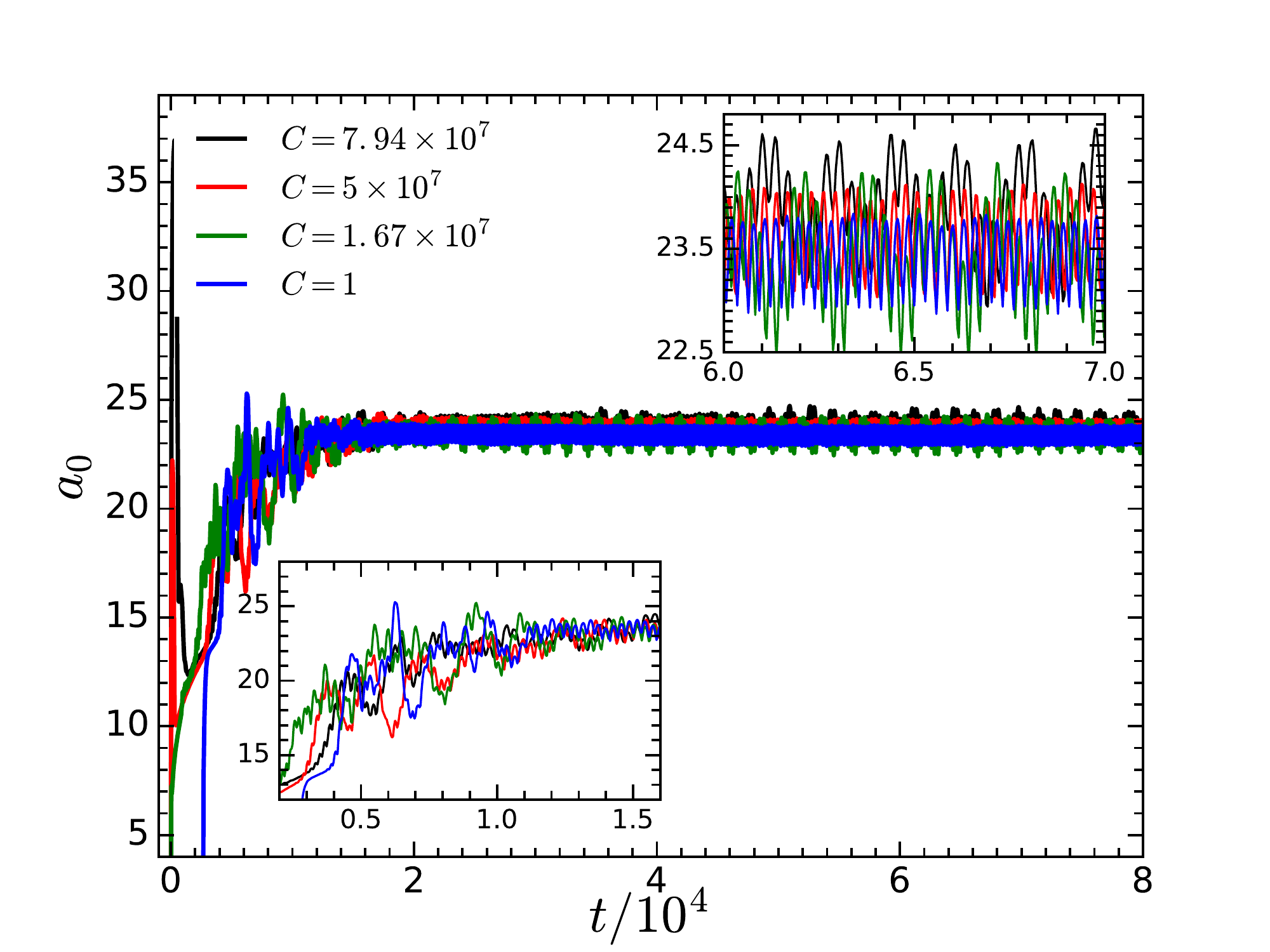}
    \caption{Temporal evolution of the amplitude $a_0$ (mean shock radius) of $l=0$ mode for different initial strength of magnetic fields and $v_{\rm in}=0.05c$.
  Zoomed in view of the initial  phase is shown in inset figure on bottom left. The inset figure on the top right shows the zoomed in view of the 
quasi-steady phase. Shock radius increases with time and eventually settles into an oscillating value. The mean shock radius is larger for a stronger field
(see inset in the top right)}
      \label{fig:amp_0}
    \end{figure}

FIG. \ref{fig:mode_c_5e7} shows the  evolution of mode amplitudes $a_l$ (measured by decomposing the shock radius into spherical harmonics; see 
Section \ref{sect:mode_analysis}) with time for the {\em Case I} MHD run.
As in HD evolution, $l=0$ is always the dominant mode.  But unlike HD, during the very early evolution of the flow ($t < 400 r_g/c$), all the 
even order modes ($l=2, 4, 6$ etc.) are more dominant compared to the odd modes ($l=1,3,5$ etc.).  This can be attributed to the anisotropic nature of the 
initial transient phase of evolution, which is symmetric about $\theta=\pi/2$ (see snapshot at $t=115.77 r_g/c$ in FIG. \ref{fig:be2e4_rho}).
When the post-shock cavity attains an almost spherical shape and starts expanding radially, contribution from even order modes with $l \geq 2$ becomes negligible. As the shock starts oscillating 
vertically about the equatorial plane, $l=1$ mode starts to dominate the higher order modes. In the fully nonlinear quasi-steady regime, 
apart from the $l=0$ mode, $l=1$, $l=2$ and $l=3$ are the most prominent modes. 

Comparing FIG. \ref{fig:mode_c_5e7} and FIG. \ref{fig:mode_c_1}, it is very difficult to %qualitatively 
quantitatively study the differences in modal contribution of the HD and MHD runs. FIG. \ref{fig:amp_0} shows 
the temporal evolution of $a_0$ (see Eq. (\ref{eq:al})), a measure of spherical radius of the aspherical shock, for different magnetic field strengths.
At the very early stage of evolution, a large value of $a_0$ reflects the transient phase at $t=115 r_g/c$ (see Fig. \ref{fig:be2e4_rho}). 
After the transient phase, a spherical shock emerges, the value of $a_0$ drops. After showing large fluctuations in $a_0$ in the transition phase, the shock attains a 
quasi-stationary state with $a_0$ oscillating about a mean value. 
The inset at top right shows the zoomed in view of $a_0(t)$ in steady state.
As expected, the average shock radius increases with an increase in the magnetic field strength. Even in the 
quasi-steady state, $a_0$ does not show sinusoidal variation at a single frequency. 

 \begin{figure}
   %\centering
    \includegraphics[scale=0.45]{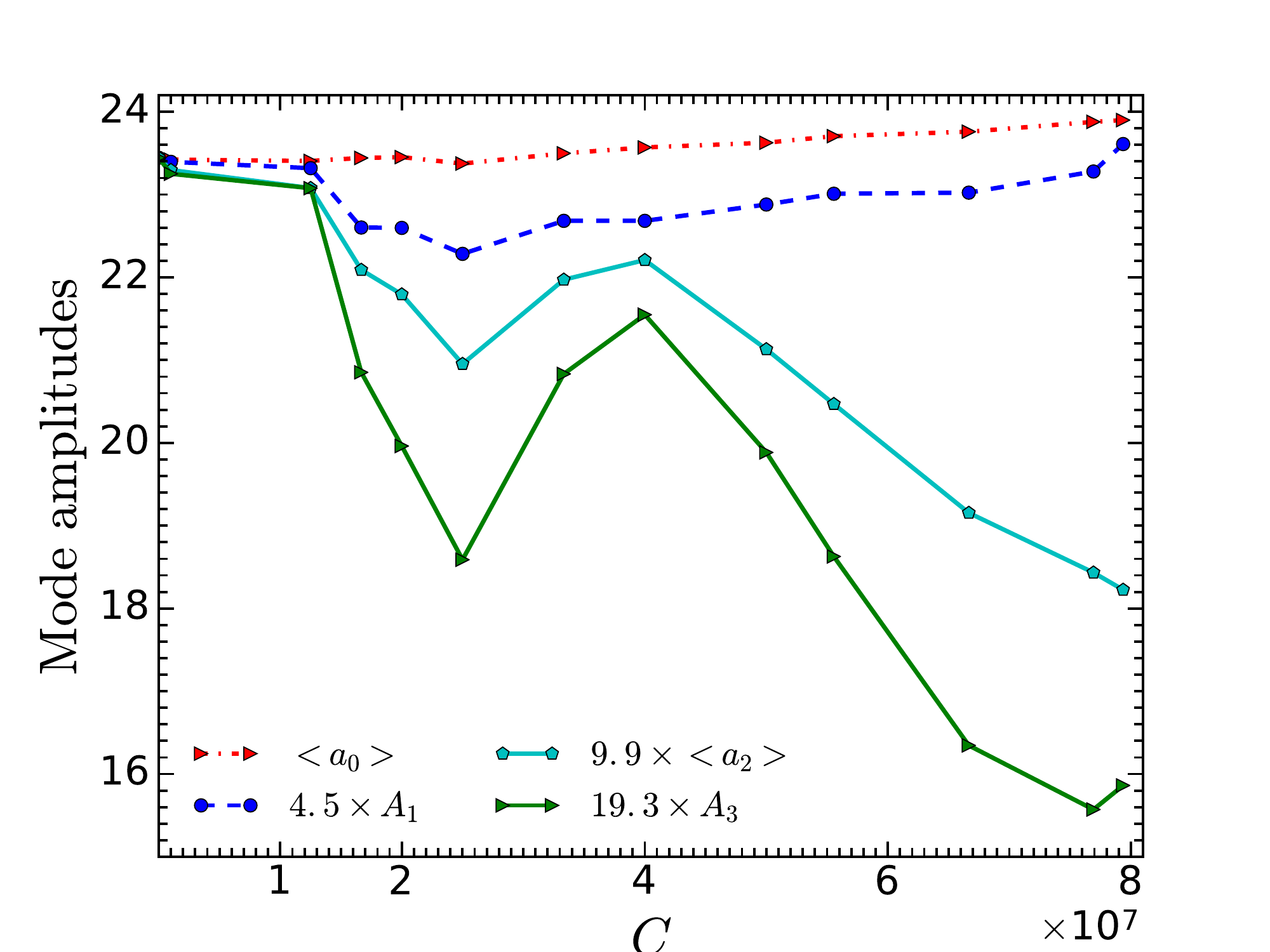}
    \caption{Variation of  mode amplitudes $a_l$ with initial magnetic field strength (quantified by $C$) for $v_{\rm in}=0.05c$. For $l=0$ and $l=2$ 
modes, amplitudes ($a_0$ and $a_2$ respectively) are time averaged in the quasi-steady state. For $l=1$ and $l=3$, the amplitudes $A_1$ and $A_3$ 
are obtained by fitting $a_1$ and $a_3$ with  fitting functions, $\psi_1 = A_1 {\rm sin}(\omega_1 t + \phi_1)$ and
 $\psi_3 = A_3 {\rm sin}(\omega_3 t + \phi_3)$ respectively.}
    \label{fig:mode_mag}
 \end{figure}
 
 FIG. \ref{fig:mode_mag} shows the variation of time-averaged mode amplitude for $l=0,~1,~2,~3$ with the initial magnetic field strength. For $a_0$ and $a_2$ we
 we take time average in the quasi-steady state. For $l=1$ and $l=3$,  as $a_1$  and $a_3$ oscillate about a vanishing mean value,  we fit the quasi-steady 
 $a_1(t)$  and $a_3(t)$ with a sinusoidal curve, and plot the variation of the amplitudes $A_1$ and $A_3$  with the magnetic field strength.  
As expected, FIG. \ref{fig:mode_mag} shows that a stronger magnetic field suppresses the higher order modes due to higher magnetic tension.

 \subsubsection{Timescales from linear theory}
 \label{sect:time_scales_MHD}
  
 Any disturbance in HD is carried by either sound waves propagating at $\pm c_s$ relative to the inflow or by entropy/vorticity waves traveling at the local flow velocity. 
 In MHD the sound wave generalizes to the fast mode and the entropy mode still consists of perturbations in total (thermal+magnetic) pressure balance.
 However, there are two new modes in MHD: the shear Alfv\'en wave and the slow magnetosonic waves.  
 Therefore, the advective part of the advective-acoustic cycle is expected to
split into five different cycles: an entropy wave, two Alfv\'en waves and two slow magnetosonic waves (\citealt{Guilet2010}). 

To interpret the SASI oscillation timescales in presence of magnetic fields, %in addition to the three timescales mentioned in section \ref{sect:time_scales_HD}
 we compute two more timescales in addition to the three timescales introduced in Section \ref{sect:time_scales_HD}.
 For computing timescales in the MHD set-up, we assume that for radial propagation (wave-vector ${\bf k} || {\bf B_0}$; fields are roughly 
 radial even in the quasi-steady state as seen in FIG. \ref{fig:be2e4_rho}), $v_{\rm slow} \approx v_A$ and $v_{\rm fast} \approx c_s$ (valid for $c_s > v_A$; see Eq. 19
 in Chapter 5 of \citealt{Kulsrud2005}; 
 FIG. \ref{fig:beta_snapshots} indeed shows that $\beta \gg 1$ throughout), and for meridional 
 propagation (${\bf k} \perp {\bf B_0}$), $v_{\rm fast} \approx \sqrt{c^2_s + v^2_A}$.

We compute two Alfv\'en/slow-magnetosonic timescales: one in which the inward propagating disturbances are propagating at the sum of local Alfv\'en and flow speeds
 \be
 \label{eq:taac_A_plus}
 t_{\rm aacA+} = \int_{r_{\rm in}}^{r_{\rm sh,I}} \frac{dr}{(\bar{c}_s(r) - |\bar{v}_r(r)|)} + \int_{r_{\rm in}}^{r_{\rm sh,O}} \frac{dr}{(|\bar{v}_{r}(r)| + \bar{v}_A(r))},
 \ee
 and $ t_{\rm aacA-}$ in which Alfv\'enic disturbances travel outward with respect to the inflow
 \be
 \label{eq:taac_A_minus}
 t_{{\rm aacA-}} = \int_{r_{\rm in}}^{r_{\rm sh,I}} \frac{dr}{(\bar{c}_s(r) - |\bar{v}_r(r)|)} + \int_{r_{\rm in}}^{r_{\rm sh,O}} \frac{dr}{(|\bar{v}_{r}(r)| - \bar{v}_A(r))}.
 \ee
 In both these cases the outward signal propagation happens at the fast speed relative to the inflow.

\begin{figure}
   %\centering
    \includegraphics[scale=0.45]{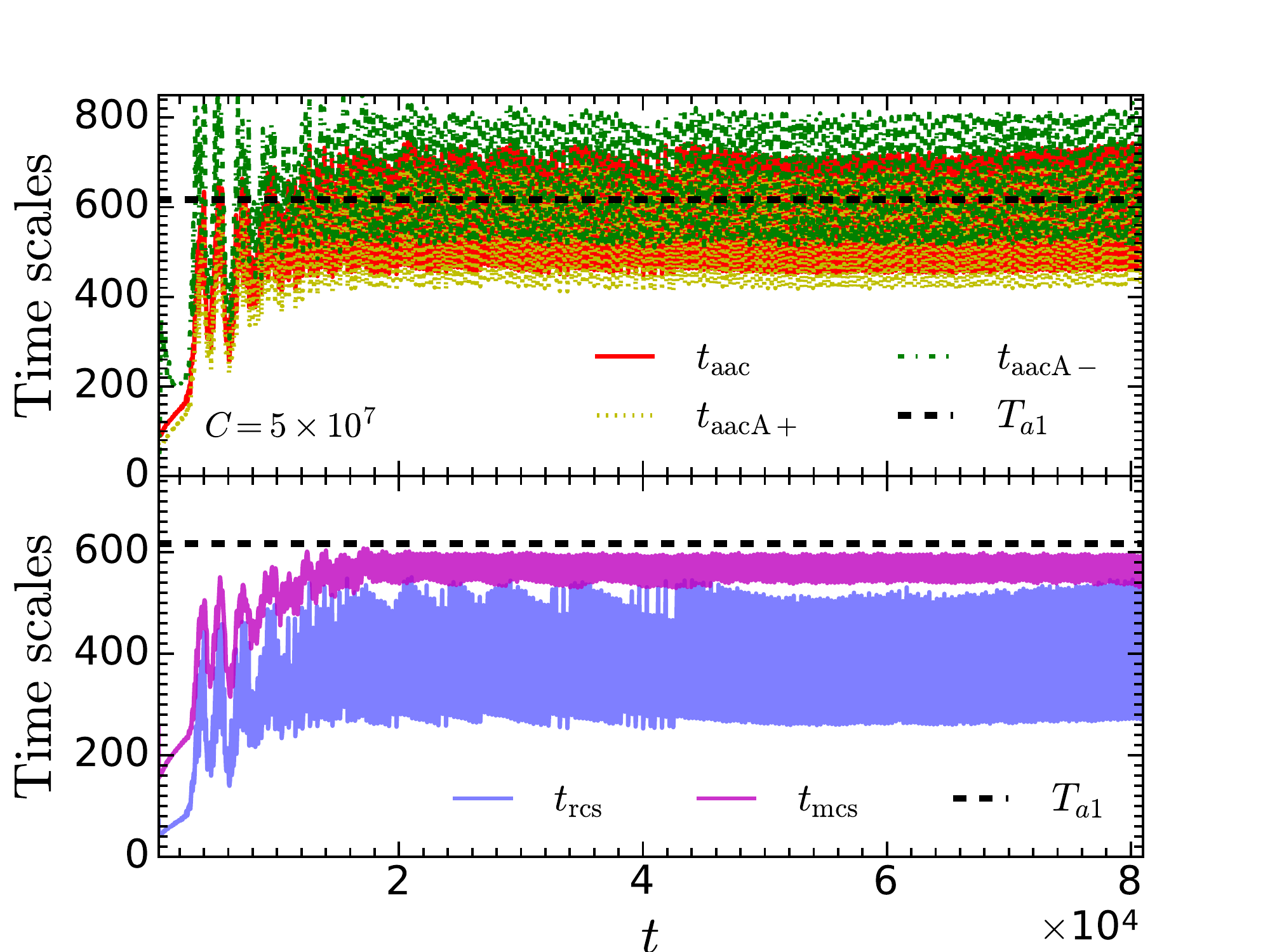}
    \caption{Various time scales as a function of time for fiducial MHD run. The observed time period of oscillation $T_{a1}=617.01$. Time averaged value of timescales are- $<t_{\rm aac}>=565.36$, $<t_{\rm rcs}>=363.61$ $<t_{\rm mcs}>=571.29$, $<t_{\rm aacA+}>=534.40$, $<t_{\rm rcs}>=630.84$. Time scales are expressed in units of $r_g/c$.}
    \label{fig:ts_c_5e7}
 \end{figure}

 FIG. \ref{fig:ts_c_5e7} shows the temporal variation of all the relevant timescales for the fiducial MHD run ({\em Case I}). 
While  $T_{a1}=617.01$ $r_g/c$ (SASI timescale measured by the period of $l=1$ perturbation in the shock location) 
still lies in the range of variations of  $t_{\rm aac}$, $t_{\rm aacA+}$ and $t_{\rm aacA-}$ (Eqs. (\ref{eq:taac}), (\ref{eq:taac_A_minus}), (\ref{eq:taac_A_plus})) 
related to the advective-acoustic cycle, it is longer than the meridional and radial sonic timescales, $t_{\rm mcs}$ and $t_{\rm rcs}$ (Eqs. (\ref{eq:tmcs}), (\ref{eq:tcs})).

\begin{figure}
   %\centering
    \includegraphics[scale=0.45]{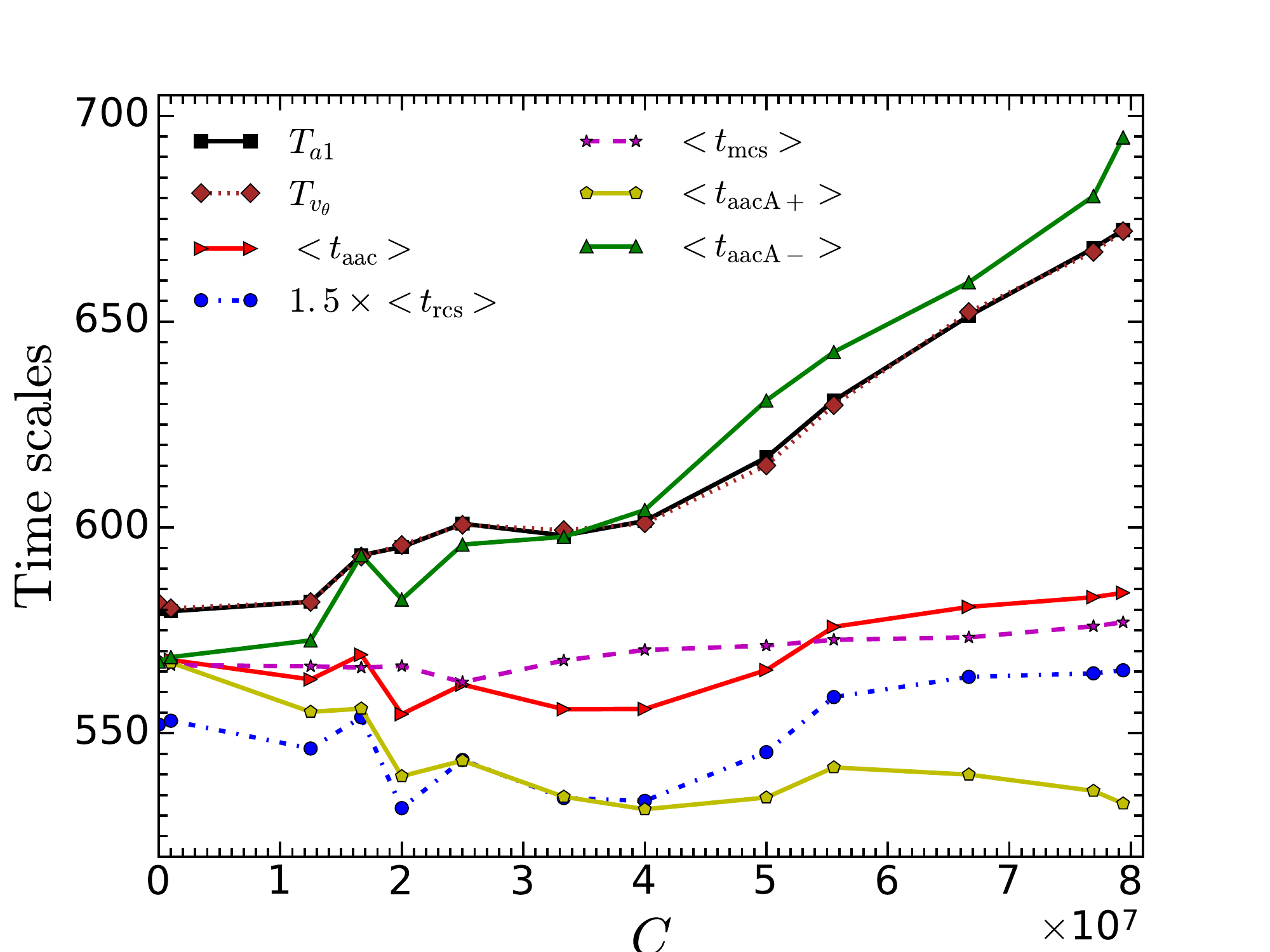}
    \caption{Variation of the observed time period of oscillation ($T_{a1}$ and $T_{v_{\theta}}$) and average (over time in quasi-steady state) 
    signal propagation timescales ($t_{\rm aac}$, $t_{\rm rcs}$, $t_{\rm mcs}$, $t_{\rm aacA+}$, $t_{\rm aacA-}$; Eqs. (\ref{eq:taac}), (\ref{eq:tcs}), 
    (\ref{eq:tmcs}), (\ref{eq:taac_A_plus}), (\ref{eq:taac_A_minus})) as a function of the initial magnetic field strength ($C$) for $v_{\rm in}=0.05c$. Time scales are expressed in units of $r_g/c$.}
    \label{fig:ts_mag}
 \end{figure}

FIG. \ref{fig:ts_mag} shows the variation of SASI time period (measured by both methods, $T_{a1}$ and $T_{v_{\theta}}$; see Section \ref{sect:sasi_period_measure}) 
and the time-averaged timescales (obtained form various signal propagation timescales) as a function of the initial magnetic field strength 
(quantified by $C$; see Eq. (\ref{eq:mag_field})). While the maximum relative change (compared to HD) in SASI time period is $\approx 15.84 \% $, 
that in $t_{\rm mcs}$ is only $\approx 1.75 \%$. In presence of a weak magnetic field ($\beta \gg 1$), SASI time period is not expected to 
be affected if the mechanism is purely acoustic. Even the variation in the HD advective-acoustic timescale
is small. However, the $t_{\rm aacA-}$ timescale in which the inward-propagating signal travels at $|\bar{v}_r|-\bar{v}_A$ (i.e., Alfv\'en wave travels outwards 
relative to the flow) matches the variation of the observed SASI timescale fairly well. In principle, the inward propagating signal should consist of five waves
 (\citealt{Guilet2010}), but a cycle consisting of outward propagating fast waves and inward-propagating Alfv\'en disturbances (traveling inward at $|\bar{v}_r|-\bar{v}_A$)
 seems to quantitatively describe the shock oscillations observed in our simulations.

 % ..............................................................................................Discussions and Conclusions ...................................................................................%
 \section{Discussions and Conclusions}
 \label{sect:discussions}
In this paper, we describe the effects of an initial split-monopolar magnetic field on the {\em standing accretion shock instability} (SASI). 
Now, we discuss the key results of our work and draw conclusions.
  
\subsection{Flow structure}
 \label{sect:flow_struct}
  In Section \ref{sect:analytic_bondi} we showed that a radial magnetic field does not modify the Bondi accretion solution. 
  In this section, we discuss how the flow structure changes in the saturated state for $v_{\rm in}=0.05c$ and different magnetic field strengths.
  Beyond a certain magnetic field  strength (for $v_{\rm in}=0.05c$, the critical value is $C=7.94 \times 10^7$)
  SASI does not occur. We choose four different strength of magnetic field: $C=1$, unmagnetized; $C=5 \times 10^7$, moderately magnetized;
  $C=7.94 \times 10^7$, the strongest magnetic field for which SASI occurs and $C=8 \times 10^7$, the strength of magnetic field at which 
  SASI can not occur.
  
\begin{figure}
    \centering
    \includegraphics[scale=0.45]{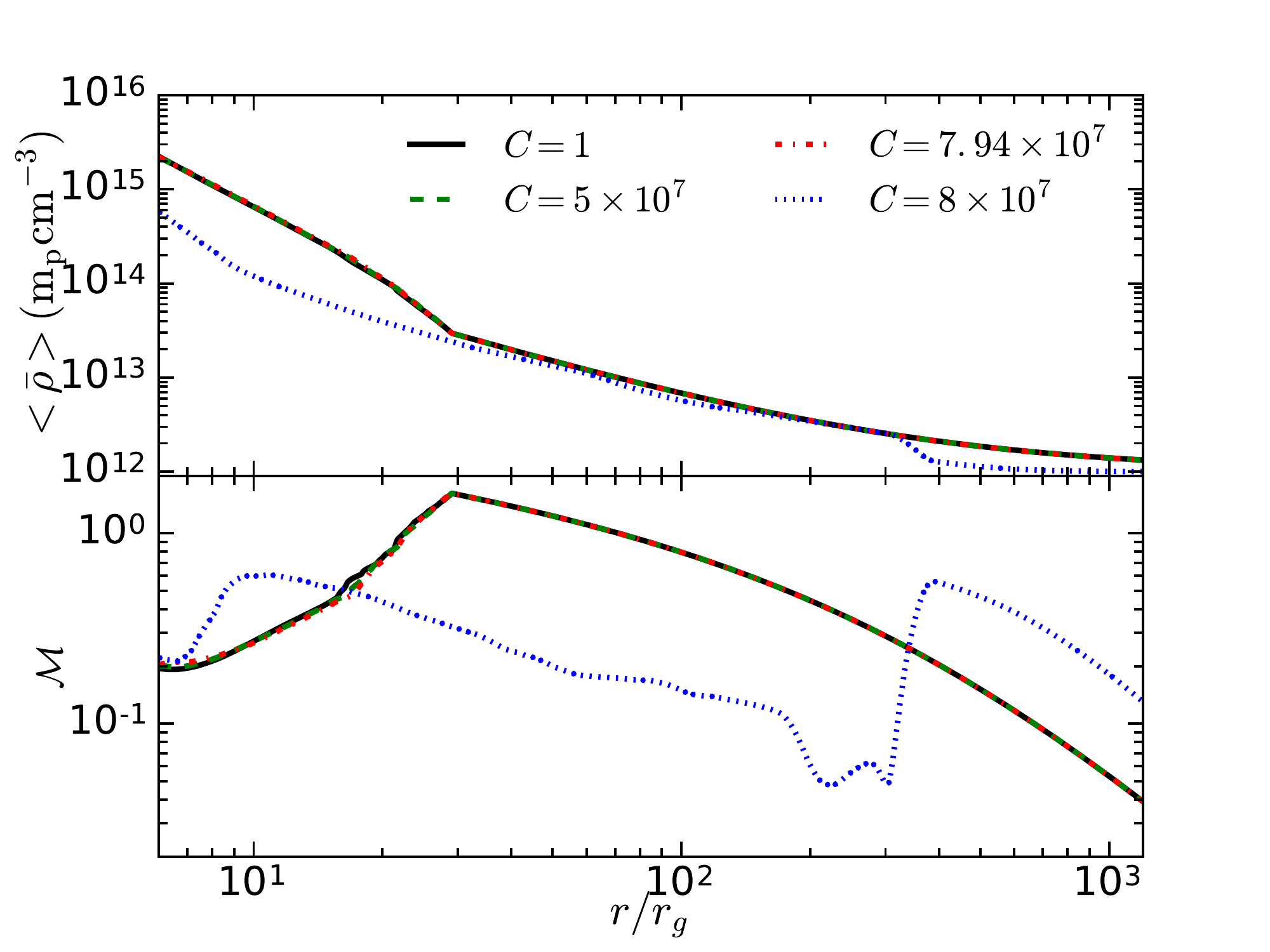}
    \caption{ Top panel: $\theta$- and time averaged density profile for different initial magnetic field strengths - $C=1$ (almost unmagnetized), $C=5 \times 10^7$
     (moderately magnetized), $C=7.94 \times 10^7$ (the strongest magnetic field for which SASI can occur) and $C=8 \times 10^7$ (magnetic field strength at 
     which SASI can not occur).
    Bottom panel: Mach number $\mathcal{M}= -<\bar{v}_r>/<\bar{c}_s> $ as a function of $r$ for different value of $C$.}
      \label{fig:steady_prof}
    \end{figure}

  FIG. \ref{fig:steady_prof} shows the average flow profiles for four different initial magnetic field strengths. 
  We take the time and angle average of the quantity $q$ as,
   \be
      <\bar{q}> = \frac{1}{T} \frac{ \int_{T_0}^{T_0+T} dt \int_{0}^{\pi}q (r, \theta, t) {\rm sin} \theta d\theta }{\int_{0}^{\pi}  {\rm sin} \theta  d\theta},
   \ee
   where $T$ is the averaging period. 
   We represent time average by `$<>$' and $\theta$- average by `$-$'. Top panel shows the  average density  as a function 
   of $r$. For magnetic field strengths which allow SASI, radial density profiles fall on top of each 
   other, irrespective of the magnetic field strength. This implies that the local flow structure may be different for different field strengths, but 
   on average the flow structures are identical. Bottom panel shows the radial profile of  local Mach number $\mathcal{M}= -<\bar{v}_r>/<\bar{c}_s>$. 
   Here also for all three magnetic field strengths for which SASI occurs, profiles are identical. But for $C= 8 \times 10^7$, for which an oscillating shock does not occur, 
   the density and Mach number profiles are different from the other three cases. The roughly hydrostatic flow is unsteady but eventually expected to asymptote to the settling 
   flow described by lower branch in Bondi solution (see the  first panel of FIG. 14 in Paper I). 
   
   Thus strong magnetic field beyond a critical strength, chokes the flow, 
   reducing the effective $v_{\rm in}$. In our idealized model, for the same sonic radius $r_c$, the critical magnetic field strength depends on the advection
   velocity at the inner boundary $v_{\rm in}$ which determines the shock radius. Larger the $v_{\rm in}$, higher the advection of thermal and magnetic energies through the 
   inner boundary.  As a result, gravity can counter stronger magnetic pressure (within post-shock cavity) which acts outward along with the thermal pressure.
 
 \subsection{SASI mechanism}
 
 \begin{figure}
   %\centering
    \includegraphics[scale=0.45]{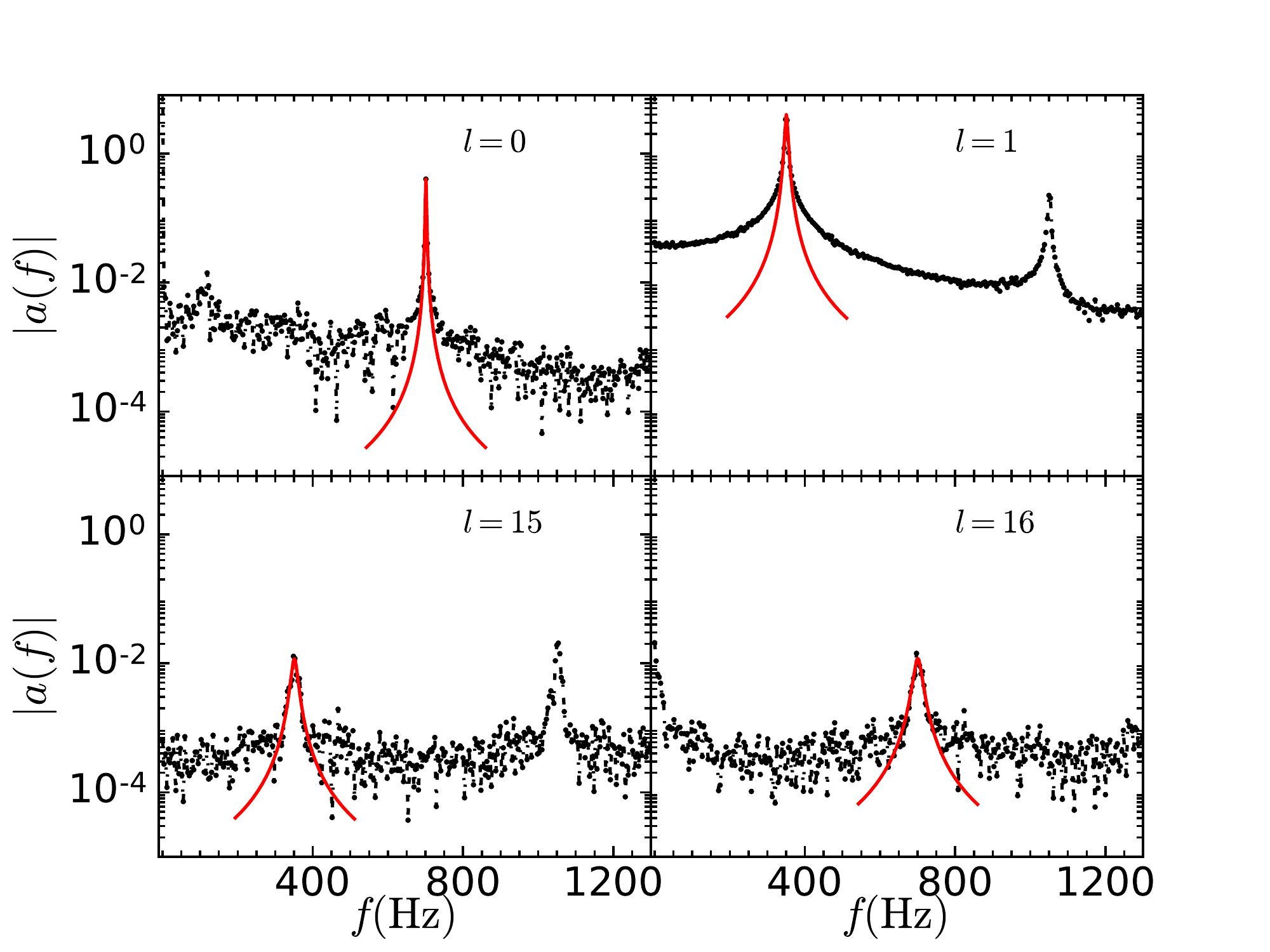}
    \caption{FFT of the mode amplitudes $a_l$ for $l=0,1,15,16$ for the fiducial hydro run. The lowest frequency peaks are fitted individually by a Lorentzian
    using the least squares fit method. Centroid of the Lorentzian gives the frequency associated with the corresponding mode.}
    \label{fig:mode_freq_hd}
 \end{figure}
 
 \begin{figure}
   %\centering
    \includegraphics[scale=0.45]{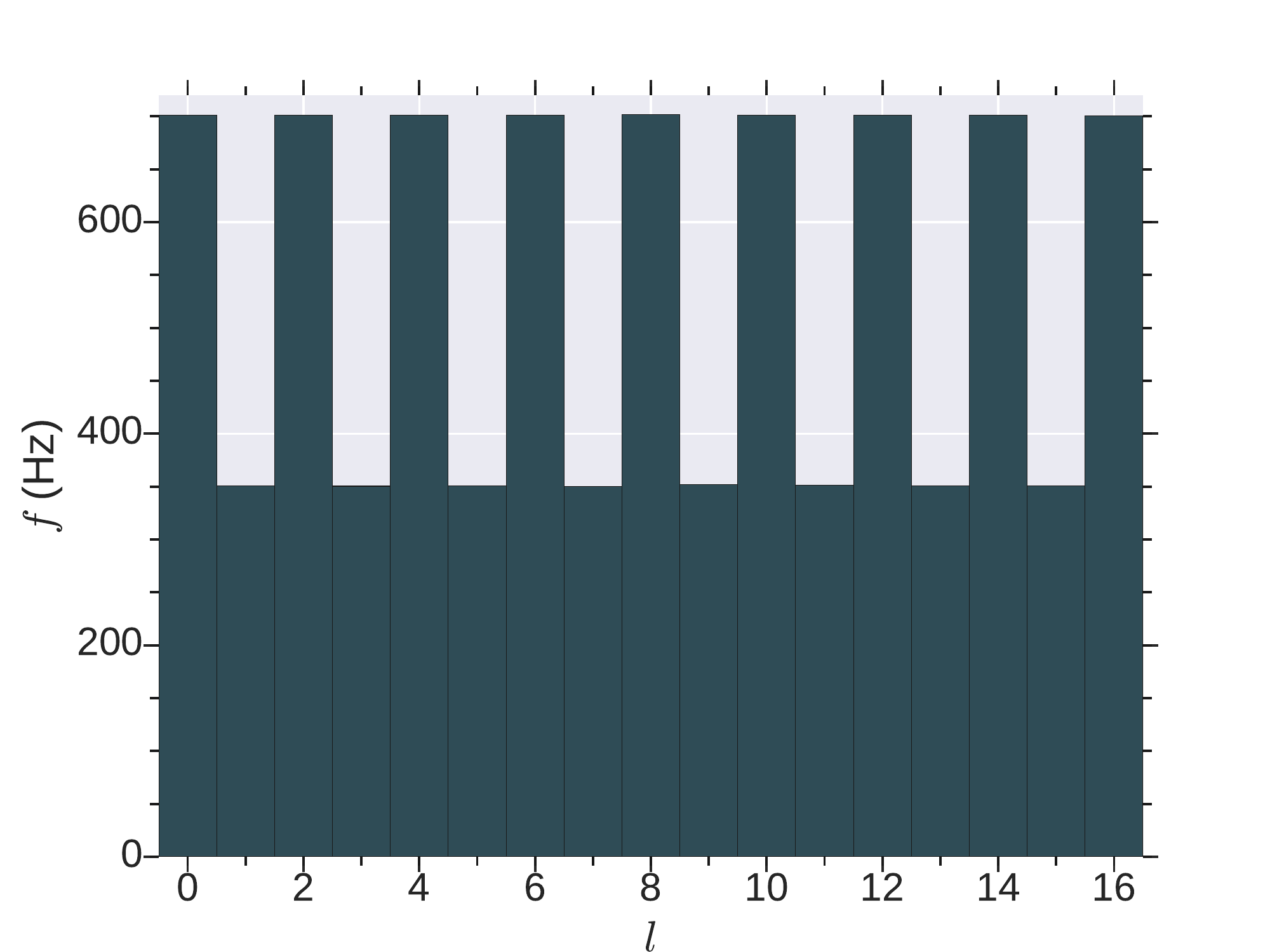}
    \caption{Lowest frequency associated with different modes ($l$) of oscillation for the fiducial HD run.}
    \label{fig:freq_mode_05}
 \end{figure}
 
  \begin{figure}
   %\centering
    \includegraphics[scale=0.45]{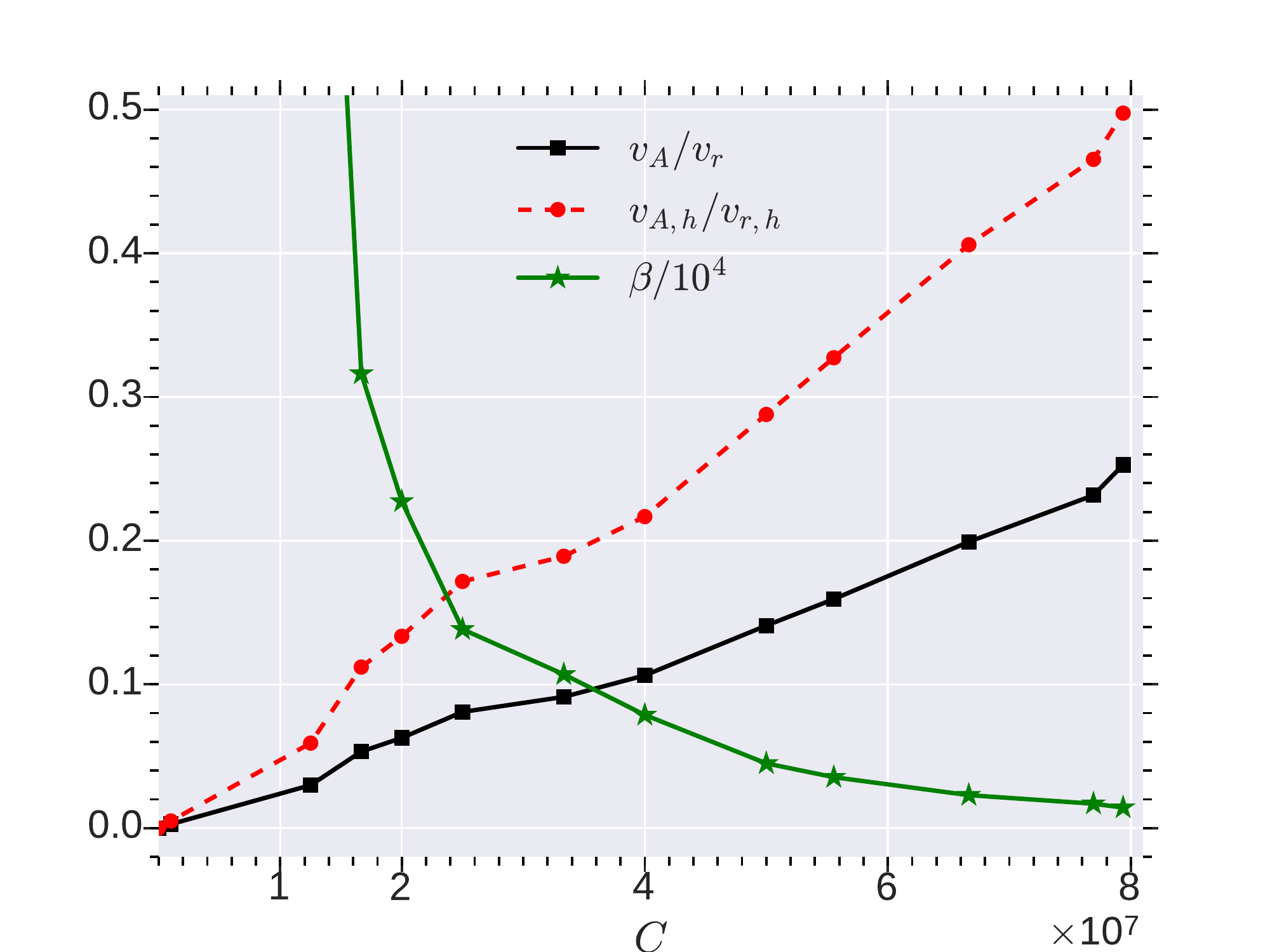}
    \caption{The ratio of average Alfv\'en speed ($v_A$) to average advection speed ($v_r$) and ratio of gas pressure to magnetic 
    pressure ($\beta$) with initial magnetic field strength ($C$). For $v_{A}/v_r$ and $\beta$ average is done over whole post-shock volume, whereas
    for $v_{A,h}/v_{r,h}$, average is done within the half shock radius. }
    \label{fig:va_vr}
 \end{figure}
 Unlike most of the previous simulations, our set-up leads to a quasi-steady state in which the nonlinear oscillations essentially
 last forever. We compare the the SASI time period with timescales obtained from two different mechanisms (namely advective-acoustic and 
 purely acoustic) in HD and MHD.
 
 In HD regime, we change the ratio of mean shock radius ($r_{\rm sh}$) to inner radius ($r_{\rm in}$) keeping $r_{\rm in}$ fixed,
 and study the variation of different timescales with this ratio. For small values of $r_{\rm sh}/r_{\rm in}$, the match between 
 the advective-acoustic timescale $t_{\rm aac}$ and the SASI time period ($T_{v_{\theta}}$ or $T_{a1}$) is excellent. With an increase
 in $r_{\rm sh}/r_{\rm in}$, the deviation of  the time scale becomes larger (see FIG. \ref{fig:rsh_rin_time}). Purely acoustic mechanism 
 gives rise to two different timescales, the radial acoustic time $t_{\rm rcs}$ (considering purely
 radial propagation) and the meridional acoustic time $t_{\rm mcs}$ (considering meridional propagation). In all cases, $t_{\rm rcs}$ is always much less
 than the SASI time period, so we can discard purely radial acoustic mechanism as the possible cause for SASI. The match between the $t_{\rm mcs}$ and 
 SASI time period becomes best around $r_{\rm sh}/r_{\rm in} \sim 3.9$. But
 it is to be noted that according to \cite{Blondin_mezzacappa2006}, SASI time period is expected to be equal to the round trip time of 
 two sound waves advancing along the shock surface i.e. $2t_{\rm mcs}$. Instead, we observe the closeness between $t_{\rm mcs}$ and SASI time 
 period.
 
 In MHD regime, we study the variation of  SASI timescales with the change in initial magnetic field strength. In presence of 
 a weak magnetic field, the advective-acoustic cycle is expected to split into five different cycles which constitute the actual cycle (see Section \ref{sect:time_scales_MHD}).
 We compute three timescales to quantify five cycles, $t_{\rm aac}$ -- outgoing fast magnetosonic wave + ingoing entropy wave, $t_{\rm aacA+}$ -- 
 outgoing fast magnetosonic wave + ingoing (with respect to local inflow) Alfv\'en/slow wave, 
 $t_{\rm aacA-}$ -- outgoing fast magnetosonic wave + outgoing (with respect to local inflow) Alfv\'en/slow wave. While the 
 maximum relative change (to HD) in timescales obtained from advective-acoustic mechanism is $ \approx 18 \%$ (for $t_{\rm aacA-}$), that in meridional acoustic 
 mechanism is only $\approx 1.75 \%$ (for $t_{\rm mcs}$);
 compared to change in SASI time period is $\approx 15.84 \%$. In purely acoustic mechanism, weak magnetic fields do not affect
 the SASI time period, but in advective-acoustic mechanism weak magnetic fields affect the SASI time period. The effects depend
 on the ratio $v_{A}/v_{r}$, the ratio of Alfv\'en and radial advection speeds.

 Both in HD and MHD regimes, advective-acoustic mechanism is favored as the possible mechanism  for SASI.
 Further, if SASI is a purely acoustic mechanism, the dispersion relation in the local limit  is given by $\omega = k c_s$, which means that
 the frequency of different modes should be proportional to the  wave number. To find the frequency associated with different modes, we take the
 temporal FFT of shock deformation modes ($a_l(t)$) and best fit the lowest frequency peak
 with a Lorentzian (Eq. (\ref{eq:Lorentzian}));  centroid frequency gives the frequency of the corresponding mode. FIG. \ref{fig:mode_freq_hd} shows the representative 
 examples of FFT  of $a_{l}$ for $l=0,1,15,16$ for our fiducial hydro run.   
 Bar plot in FIG. \ref{fig:freq_mode_05} shows the variation of  mode frequency with mode number. All the even modes ($l=0,2,4$ etc.) have frequency 
 $\approx 700$ Hz, and odd modes ($l=1,3,5$ etc.) have frequency $\approx 350$ Hz, which is against the expectation $f \propto l$ for acoustic signals.
 
 \cite{Guilet2010}, in their toy model, showed that 
 the effects of magnetic field on the advective-acoustic cycle depend on the ratio of Alfv\'en speed to advection speed ($v_A/v_r$) instead
 of the ratio of gas pressure to magnetic pressure ($\beta$). Even a weak magnetic field is able to significantly affect the advective-acoustic cycle
 provided $v_A/v_r$ is of the order of
 \be
 \label{eq:ratio}
 \frac{v_A}{v_r} \sim \frac{r_{\rm sh}}{2h\sqrt{l(l+1)}}  ,
 \ee
 where $r_{\rm sh}$ is the shock radius, $h$ is the distance over which flow is decelerated, and $l$ is mode number (see their Eq. 18). 
 If we take $r_{\rm sh} \approx h$, we get the ratio $0.353$ and $0.204$ for $l=1$ and $l=2$ modes respectively. 
 FIG. \ref{fig:va_vr} shows the change in ratio of Alfv\'en speed to advection speed considering the average over the whole post-shock volume ($v_A/v_r$) 
 and over the volume within half shock radius ($v_{A,h}/v_{r,h}$) with the change in the initial magnetic field strength ($C$). To quantify the magnetization of 
 the accreting medium in the quasi-steady state, the ratio of  average gas pressure to average magnetic pressure in the post-shock volume ($\beta$) is 
 plotted as a function of $C$ in FIG. \ref{fig:va_vr}. SASI time period ($T_{a1}$, $T_{v_{\theta}}$)
 and timescales corresponding to the advective-acoustic mechanism ($t_{\rm aac}$, $t_{\rm aacA+}$ and $t_{\rm aacA-}$) encounter 
 a significant change from the  timescales in hydrodynamic case for $C=1.67 \times 10^7$ (as seen in FIG. \ref{fig:ts_mag}), for which $v_A/v_r = 0.053$ and
 $v_{A,h}/v_{r,h}=0.11$ and $\beta \approx 3200$. So it appears that in our set-up, SASI is affected at a smaller value of $v_{A}/v_r$ 
 compared to the estimates of \cite{Guilet2010}. This is not surprising, because we use an initial split-monopolar magnetic field 
configuration which leaves its imprint even in the quasi-steady state. Therefore, magnetic field is stronger at small $r$ in contrast to the uniform 
field distribution used by \cite{Guilet2010}. This argument is supported
by the larger value of $v_{A,h}/v_{r,h}$ (when average is done over the volume within half the shock radius) compared to $v_A/v_r$
(when average is done over whole post-shock volume) for the same value of $C$ (see FIG. \ref{fig:va_vr}). Moreover, the shape of $v_A/v_r$ versus $C$ curve 
(in FIG. \ref{fig:va_vr}) and $T_{a1}$ or $T_{v_{\theta}}$ 
versus $C$ (in FIG. \ref{fig:ts_mag}) are very similar; whenever ratio increases or decreases the observed time periods follow them. This is 
expected, because $$\int \frac{dr}{(|v_r| - v_A)} \approx \int \frac{dr}{|v_r|}\left (1+\frac{v_A}{|v_r|}\right),$$ if $v_A < v_r$.

 Eq. (\ref{eq:ratio}) tells that for the same strength of magnetic field, the higher order modes are more affected compared to lower order modes, which is clear 
 from FIG. \ref{fig:mode_mag}. We also see that the SASI period is not a monotonically increasing functions of magnetic field strength; there are irregularities 
 which are expected in the framework of advective-acoustic mechanism due to interference of different cycles (see FIG. 6 of \citealt{Guilet2010}).
 
 So we conclude that the physical mechanism behind SASI (at least in the parameter regime that we explored) is more likely to be 
 the advective-acoustic mechanism instead of a purely acoustic mechanism (either meridional or radial).

 \subsection{QPOs and SASI}
  \begin{figure*}
   %\centering
    \includegraphics[scale=0.72]{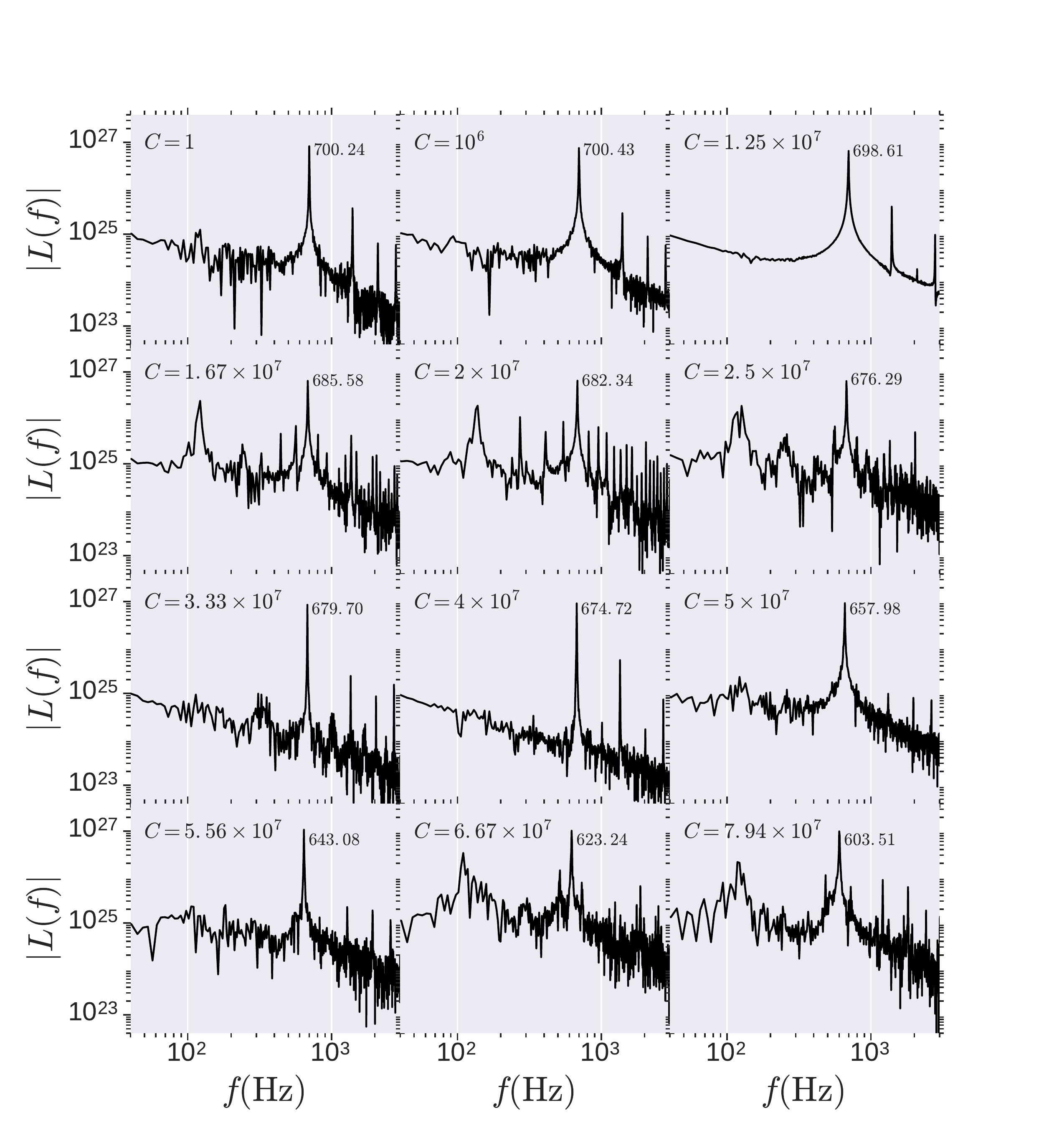}
    \caption{Power spectrum of the light curves assuming free-free emission for different magnetic field strength ($C$) and $v_{\rm in}=0.05c$. 
    With the increment in magnetic field strength 
    low frequency features appear and disappear non-monotonically. The most prominent peak whose frequency
    is the frequency of $l=0$  mode and half the frequency of $l=1$ mode can be related to upper kHz 
    QPO. For some strengths of magnetic field, alongside the main peak, there are some comparatively low
    frequency peaks which can be related to some other types of high frequency QPOs such as 
    lower kHz QPO and hHz QPO.}
    \label{fig:fft_L}
 \end{figure*}
 
 \begin{figure}
   %\centering
    \includegraphics[scale=0.35]{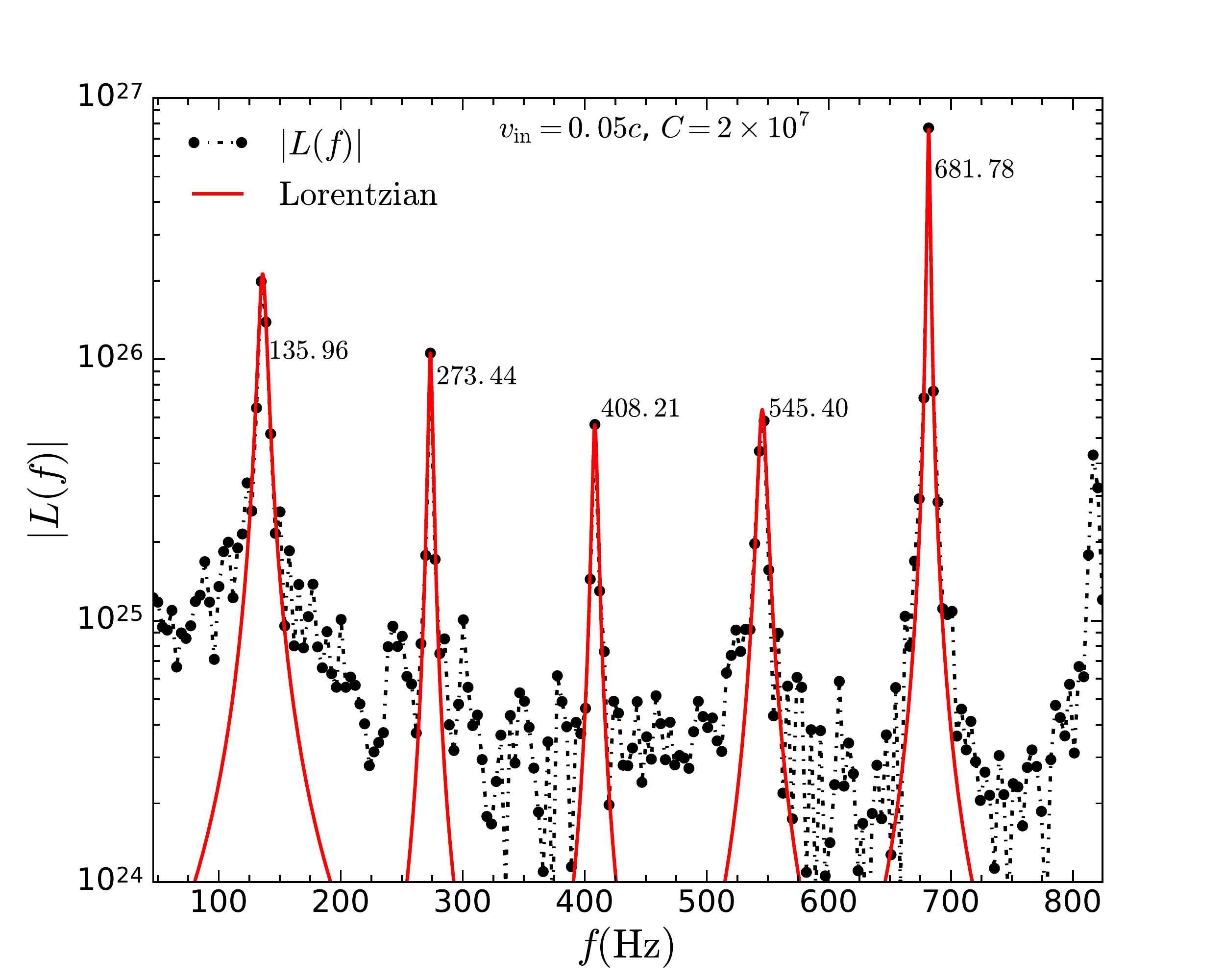}
    \caption{Power spectrum of light curve for $v_{\rm in}=0.05 c$ and $C=2 \times 10^7$. The prominent peaks appears at $f_u=681.78$ Hz,
    $f_{hHz}=135.96$ Hz, and at its harmonics, $f_{hHz2}=273.44$ Hz, and  at the beat frequencies, $f_{l1}=545.40$ Hz$\sim f_u - f_{hHz}$, 
    $f_{l2}=408.21$ Hz$\sim f_u -f_{hHz2}$. The peaks are fitted individually by Lorentzians
    using least squares fit method.}
    \label{fig:fft_lh_05}
 \end{figure}
 
 \begin{figure}
   %\centering
    \includegraphics[scale=0.35]{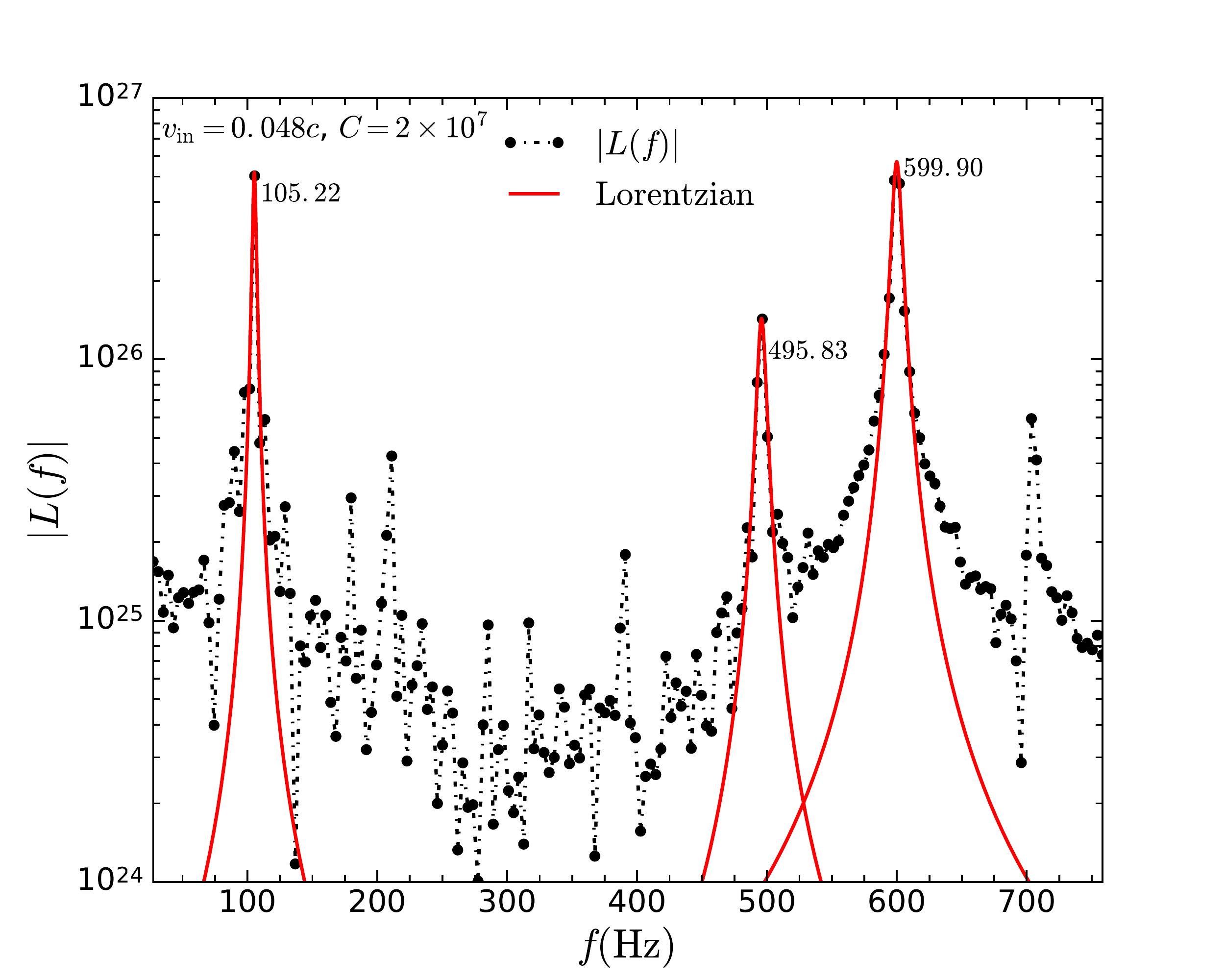}
    \caption{Power spectrum of light curve for $v_{\rm in}=0.048 c$ and $C=2 \times 10^7$. The prominent peaks appears at $f_u=599.90$ Hz,
    $f_{hHz}=105.22$ Hz, and at the beat frequency of the former two, $f_{l}=405.82$ Hz$\sim f_u - f_{hHz}$. The peaks are fitted individually by  Lorentzians
    using least squares fit method.}
    \label{fig:fft_lh_048}
 \end{figure}
 
 \begin{figure}
   %\centering
    \includegraphics[scale=0.35]{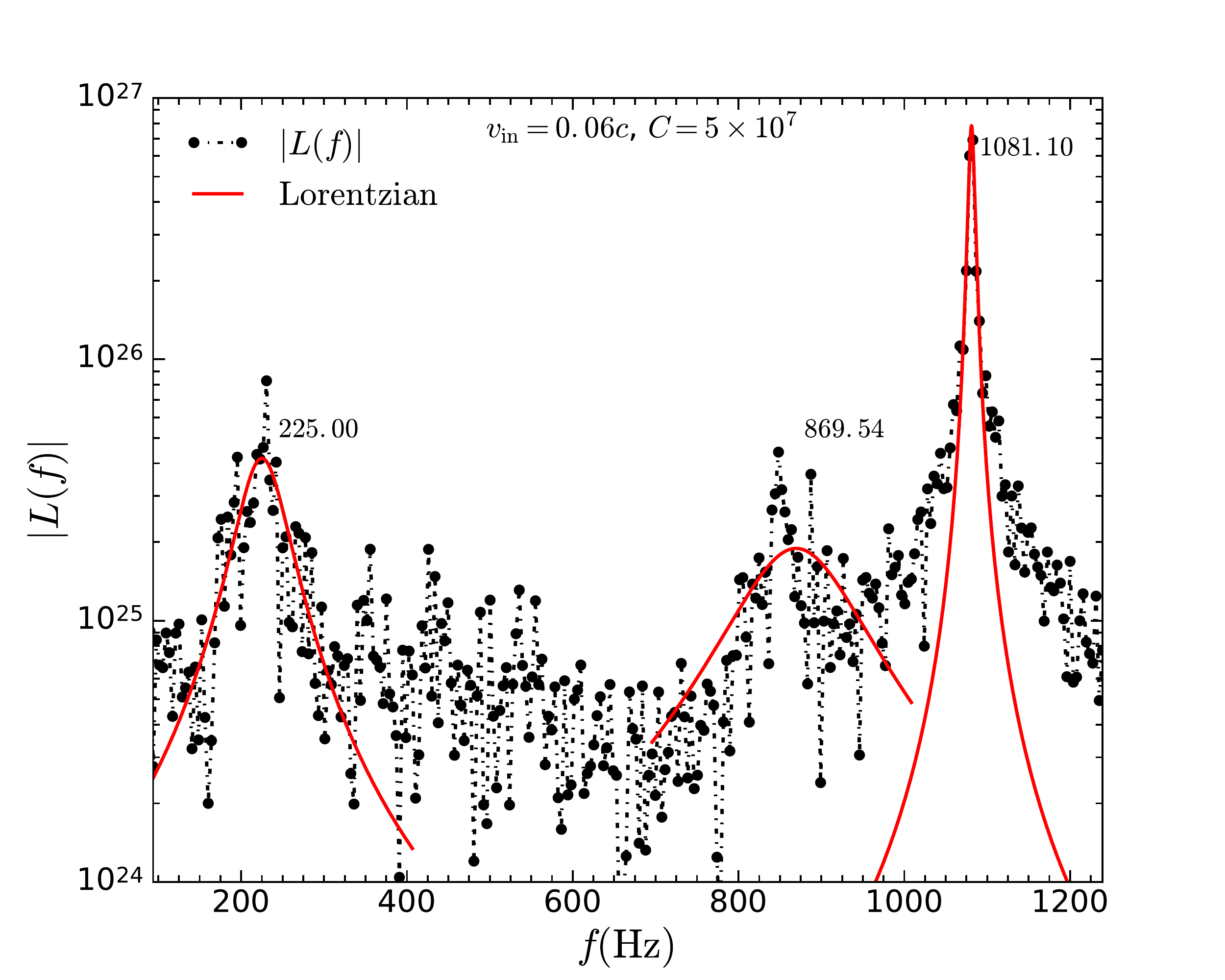}
    \caption{Power spectrum of light curve for $v_{\rm in}=0.06 c$ and $C=5 \times 10^7$. The most prominent peaks appears at $f_u=1081.10$ Hz,
    and at $f_h=225$ Hz which can be considered as upper kHz QPO and hHz QPO respectively. A fainter peak is seen at
    $f_{l}=869.54$ Hz, which can be considered as lower kHz QPO. The peaks are fitted individually by Lorentzians
    using least squares fit method.}
    \label{fig:fft_lh_06}
 \end{figure}

{\em Standing accretion shock instability} (SASI) in an accretion flow gives rise to an intrinsic time variability in the flow, which 
may explain some of the quasi-periodic oscillations (QPOs) observed in X-ray binaries. In this
section we try to connect different time scales associated with magnetized SASI with different high frequency ($\gtrsim 100$ Hz) QPOs observed 
in X-ray binaries (both in black hole and neutron star binaries).

Kilohertz (kHz) QPOs are the fastest variability components in neutron star X-ray binaries (\citealt{Vanderklis2004}), seen in most 
Z and atoll sources. Sometimes kHz QPOs appear in pairs; the peak with the higher frequency is called the upper kHz QPO at 
frequency $f_u$ and the other is called lower kHz QPO with frequency $f_l$. Many models associate orbital motion in the disk with one of the 
kHz QPOs (\citealt{Strohmayer1996}, \citealt{Miller1998}, \citealt{Mukhopadhyay_QPO2003}). While \citealt{Mukhopadhyay_QPO2003} attributes global shock oscillations
as the origin of upper kHz QPOs for the first time, some other models argue that both QPOs arise via 
nonlinear resonance between fundamental frequencies, e.g., between radial and vertical epicyclic oscillation frequencies along with the spin frequency
 of neutron star (\citealt{Kluzniak2004}, \citealt{Petri2005}, \citealt{Blaes2007}, \citealt{Mukhopadhyay2009}).
Parametric resonance models are particularly attractive if $f_u-f_l$ is linked with the 
spin frequency $\nu_s$ of the neutron star, when $f_u-f_l\sim \nu_s/2$ (if $\nu_s \gtrsim 400$ Hz; e.g. KS~1731-260, 4U~1636-53) or 
$f_u-f_l\sim \nu_s$ (if $\nu_s \lesssim 400$ Hz; e.g. 4U~1728-34, 4U~1702-429) (\citealt{Strohmayer1996},  
\citealt{Vanderklis1996}, \citealt{Ford2000}, \citealt{Wijnands2003}). However, later on this 
interpretation was questioned (\citealt{Mendez2007}).

For black hole sources, on the other hand, the observed twin high frequency (HF) 
QPOs are often argued to be seen in $2:3$ ratio [e.g GRO~J1655-40 ($300,~450$ Hz; \citealt{Remillard1999}; \citealt{Strohmayer2001a}),
XTE~J1550-564 ($184,~276$ Hz; \citealt{Homan2001}) and GRS~1915+105 ($113,~168$ Hz; \citealt{Remillard2004})], which again was explained 
based on nonlinear resonance by the groups mentioned above. Some recent observations question the $2:3$ appearance of HF QPOs in black hole 
X-ray binaries [e.g IGR~J17091-3624 ($66, ~164$ Hz; \citealt{Altamirano2012})].

Another $\gtrsim100$ Hz variability phenomenon is the hectohertz (hHz) QPO (\citealt{Ford_vanderklis1998}) with a frequency $f_{hHz}$ in the range $100-270$ Hz 
(e.g see \citealt{Altamirano2008}) in atoll sources in most states. \cite{Fragile2001} proposed that accreting material passing through the transition region 
formed due to Bardeen-Petterson effect may generate hHz frequencies. \cite{Kato2007} proposed that a warp in accretion disk gives rise to the hectohertz QPOs 
in atoll sources. The black hole sources also exhibit QPO frequency of order
hHz or slightly less, e.g. GRS~1915+105, XTE~J1550-564, simultaneously with high frequency ones (e.g. \citealt{Remillard2002}). Earlier nonlinear resonance 
models can be modified to explain it (\citealt{Mukhopadhyay2012}).

FIG. \ref{fig:fft_L} shows the  power spectrum of the  light curve $L(t)$ obtained in the quasi-steady state for different initial magnetic 
field strengths (quantified by $C$; see Eq. \ref{eq:mag_field}) and $v_{\rm in}=0.05c$.
Luminosity  $L(t)$ is assumed to be due to free-free emission (a similar time variability is expected for other mechanisms such as synchrotron and inverse-Compton) from the volume $V$, and computed as,
\be
L  = \int_{V} 1.4 \times 10^{-27} \left(\frac{\rho}{\rm m_p} \right )^2 T^{\frac{1}{2}} dV ~ {\rm erg ~s}^{-1},
\ee
where, $V$ is the spherical volume of radius $r=30 r_g$, dominated by post-shock region.
The post-shock temperature in simulations is very high ($T \sim 10^{11} K$).
The electrons in hot accretion flows are at lower temperature compared to that of the ions and other emission process
may be important (e.g. \citealt{Sharma2007}; \citealt{Rajesh2010}, \citealt{Yuan2014}). Therefore light curves 
from simulations (which assume a single temperature fluid) should be only taken as trends.

First panel of FIG. \ref{fig:fft_L} shows the power spectrum for the unmagnetized ($C=1$) SASI run and the power spectrum has the 
most prominent peak at $f_0=700.24$ Hz along with its harmonics (the peak frequency is obtained by fitting with a Lorentzian).
This is the frequency associated with the $l=0$ mode and double the frequency of $l=1$ mode. There is some low frequency
noise present in the power spectrum. With the increase in magnetic field strength, the prominent peak frequency shifts to lower value and
the low frequency noise becomes less prominent (for $C=10^6$) and vanishes for $C=1.25 \times 10^7$. As the magnetic field strength is increased 
more, some extra peaks  arise at low and intermediate frequencies  along with the main peak (e.g. $C=1.67 \times 10^7$ and 
$C=2 \times 10^7$). The lowest frequency is associated with the modulation frequency on top of a regular frequency of mode amplitude $a_0(t)$ 
(e.g. see the variation of $a_0(t)$ for $C=7.94 \times 10^7$ in FIG. \ref{fig:amp_0}). With the increase in magnetic field strength, low frequency 
features appear and disappear non-monotonically.

In the present analysis, the origin of QPOs (whether kHz, HF or hHz) is different from past proposals.
FIG \ref{fig:fft_lh_05} shows the power spectrum for $C=2 \times 10^7$ and $v_{\rm in}=0.05c$. 
The most prominent peak appears at $681.78$ Hz, which can be related to the upper kHz QPOs at $f_u$. The lowest frequency peak is at $135.96$ Hz,
which can be identified as the hHz QPO at $f_{hHz}$. In between these two peaks, there are three more peaks.
One of them is the harmonic of the hHz QPO ($f_{hHz2} = 273.44$ Hz $\approx 2f_{hHz}$), other two peaks are the beat 
frequencies, $f_{l1}=545.40$ Hz $\approx f_u-f_{hHz}$, and $f_{l2} = 408.21$ Hz $\approx f_u-f_{hHz2}$, one of which 
can be related to the lower kHz QPO. Whereas, $f_u$ is equal to the frequency associated with $l=0$  mode, $f_{hHz}$ is the  frequency of modulation 
in the mode amplitude $a_0$ due to magnetic field, as seen in inset at upper right 
of FIG. \ref{fig:amp_0}. With magnetic field strength ($C$), the upper kHz QPO frequency $f_u$ tracks $2/T_{a1}$, the frequency of $l=0$ mode
(which is double the $l=1$ mode frequency). For all field strengths, the hHz QPO frequency remains constrained in the range ($110-135$) Hz.

If the shock location is changed by tuning the value of $v_{\rm in}$, SASI time period changes (as shown in FIG. \ref{fig:rsh_rin_time}),
so does $f_u$, as $f_u \approx 2/T_{a1}$. On the other hand, the hHz QPO arises due to magnetic effects. To see
the variation in $f_{hHz}$ with the change in shock location we decrease and increase the shock radius by changing $v_{\rm in}$ to $0.06c$ and
$0.048c$ respectively. For $v_{\rm in}=0.048 c$, power spectrum of the  light curve for $C=2 \times 10^7$ is shown in FIG. \ref{fig:fft_lh_048}; 
three peaks are present in the power spectrum. Upper kHz QPO frequency $f_u=599.90$ Hz is shifted to a lower value compared to the 
fiducial case ($v_{\rm in}=0.05c$), so does the hHz QPO 
frequency ($f_{hHz}=105.22$ Hz). The frequency of lower kHz QPO is $f_l=495.82$ Hz $\approx f_u - f_{hHz}$.
FIG. \ref{fig:fft_lh_06} shows the power spectrum of light curve  for $v_{\rm in}=0.06c$ and $C=5 \times 10^7$, with a smaller shock radius.  
As expected, the upper kHz QPO frequency ($f_u=1081.10$ Hz) related to SASI time period, increases. Also, the hHz QPO frequency ($f_{hHz}=225$ Hz) moves
to a higher value. The lower kHz QPO ($f_l=869.54$ Hz) structure becomes fainter (this might be immersed in noise in real observations). 

Our idealized simulations suggest shock oscillations as the  origin of QPOs. in particular kHz/HF/hHz ones.  We identify the $l=0$ SASI mode 
frequency (which is double the frequency of $l=1$ mode)  as the frequency of upper kHz QPO.
It is the appearance of the hHz QPO which determines the separation of twin QPO peaks. We do not observe the hHZ QPOs
in our simulations without magnetic fields, indicating that they originate only in the presence of a magnetic field. Hence, one does not necessarily need to
introduce the spin of the compact objects to explain QPOs. 

\subsection{Caveats of the model}
The present model is very simplistic. In reality, accretion flows have complicated magnetic field geometry, angular momentum, cooling 
(depending on the spectral state of the XRBs) which might change the above results. A brief discussion of the above-mentioned factors is given below.

We initialize the simplest magnetic field configuration, a split-monopole. Because 
of the absence of magnetic force in the pre-shock flow, the equilibrium is not affected by the magnetic field 
(see Section \ref{sect:analytic_bondi}), but the mode frequencies are. The mode frequencies are expected to behave differently for different field geometries (\citealt{Guilet2010}). 

Accreting matter in XRBs is expected to have angular momentum. QPOs are observed in the hard state, in which the inner flow is expected to be hot, 
quasi-spherical and sub-Keplerian (e.g see \citealt{Chakrabarti1989}). To approximate  that we study the spherical, adiabatic, non-rotating accretion flow on to a compact object. However, even small angular momentum can affect the global shock oscillations (\citealt{Blondin_mezzacappa2006}, \citealt{Yamasaki2008}, 
\citealt{Kazeroni2017}). Shock instabilities in rotating accretion flows were invoked to explain time variability (mostly low frequency phenomena) in accreting 
systems (\citealt{MSC1996}, \citealt{Nagakura2009}).

We also assume axisymmetry, breakdown of which may significantly alter the oscillation frequencies. While in the non-rotating flow,
non-axisymmetric modes of SASI redistribute angular momentum (\citealt{Blondin2007}, \citealt{Fernandez2010}, \citealt{Guilet2014}, \citealt{Kazeroni2016}), in a rotating
flow the spiral modes become more prominent over the sloshing modes (\citealt{Iwakami2009}, \citealt{Kazeroni2017}) and dominate the dynamics of the flow. The occurrence of non-axisymmetric Papaloizou-Pringle instability (\citealt{Papaloizou_pringle1984}), its interplay with the advective-acoustic cycle 
(\citealt{Gu2003}), and the magneto-rotational instability (\citealt{Balbus_hawley1991}) are additional complications in a rotating accretion flow.

We also neglect radiative cooling because in the hard state, the inner accretion flow is expected to be hot and non-radiative with the cooling time much longer 
than the infall time of the  flow. Some earlier studies showed that the shock is unstable even in 1D in presence of  cooling (\citealt{Langer1981}, \citealt{Chevalier1982}, \citealt{Saxton2002}), but our non-radiative simulations require 2D or 3D to be unstable (see Paper I).

We expect shock oscillations in inner, hot, transonic accretion flows (rotating or non-rotating). But for comparing with the observations, one needs to 
study them in more realistic 3D simulations with rotation, magnetic fields.

\section{Summary}
\label{sect:summary}
In this work we study {\em standing accretion shock instability} (SASI) in unmagnetized and magnetized spherical accretion flow
around a central gravitating accretor, in particular the ones with a hard surface. The key findings of the work are listed below.

\begin{itemize}
 \item A standing shock does not occur above a critical strength of magnetic field as the sum of outward magnetic and thermal pressure 
 becomes high enough to overcome the inward gravitational attraction and the shock moves into the subsonic region and vanishes.
 \item A comparison of various signal propagation timescales and the observed shock oscillation frequency agrees with the advective acoustic
 mechanism, and not a purely acoustic one (at least for our parameters).
 \item The global shock oscillations in the accretion flow give rise to a prominent peak in the power spectrum of the light curve
 which can be related to the upper kHz QPOs. In presence of magnetic field, there are a few low frequency
 peaks that can be related to lower kHz and hHz QPOs.
 
\end{itemize}

\section*{Acknowledgments}
We thank Srikara S (IISER Pune) for taking part in discussions during initial part  of the work.
PD thanks Nagendra Kumar for helpful discussions on QPOs. We thank the anonymous reviewer for
thoughtful suggestions. This work is partly supported by an 
India-Israel joint research grant (6-10/2014[IC]) and by the ISRO project
with research Grant No. ISTC/PPH/BMP/0362. PS and PD thank KITP for their participation in 
the program ``Confronting MHD Theories of Accretion Disks with Observations''. This research was 
supported in part by the National Science Foundation under Grant No. NSF PHY-1125915.  Some of the 
simulations were carried out on  Cray XC40-SahasraT cluster  at Supercomputing Education
and Research Centre (SERC), IISc. 
 
\bibliographystyle{mnras}
\bibliography{bibtex}
\label{lastpage}

\end{document}